\documentclass[aps,twocolumn,groupedaddress,amssymb]{revtex4}
\usepackage{setspace}

\usepackage{latexsym}
\usepackage{amsbsy}
\usepackage{amsmath}
\usepackage{graphicx}
\usepackage{xcolor}

\usepackage{nameref}  

\usepackage{hyperref} 
\usepackage[capitalise]{cleveref} 
\hypersetup{
    colorlinks=true,
    linkcolor=black,
    urlcolor=blue,
    citecolor=blue
}

\begin{document}

\author{Nazma Firdosh\textsuperscript{1}, Shreyashi Sinha\textsuperscript{1} and Sujit Manna\textsuperscript{1}\footnote{Contact author: smanna@physics.iitd.ac.in}}
\affiliation{\textsuperscript{1}Department of Physics, {Indian Institute of Technology Delhi}, Hauz Khas, New Delhi 110016, India}

\preprint{APS/123-QED}

\title{Magnetic criticality and magnetocaloric response in MnBi$_2$Te$_4$ and MnBi$_4$Te$_7$
}

\date{\today}

\begin{abstract}

MnBi$_2$Te$_4$ and MnBi$_4$Te$_7$ are antiferromagnetic topological insulators belonging to the MnBi$_{2n}$Te$_{3n+1}$ series, where structural layering provides a natural route to tune magnetic interaction in van der Waals magnets. Despite extensive interest in their topological properties, how the insertion of Bi$_2$Te$_3$ quintuple layers modifies magnetic critical fluctuations near the antiferromagnetic transition remains unresolved. Here, we combine scanning tunneling microscopy (STM), critical scaling analysis, and magnetocaloric measurements to directly correlate real-space structures with magnetic criticality. STM reveals atomically flat septuple-layer terraces in MnBi$_2$Te$_4$ whereas MnBi$_4$Te$_7$  displays coexisting septuple and quintuple layer terminations reflecting its alternating stacking sequence. MnBi$_2$Te$_4$ exhibits robust three-dimensional Ising-like critical behavior together with a distinct low-temperature first-order transition. In contrast, MnBi$_4$Te$_7$ displays crossover-dominated criticality arising from weakened interlayer exchange and competing magnetic phases. Correspondingly, the magnetocaloric response differs significantly between the two compounds. MnBi$_2$Te$_4$ shows dual-type magnetocaloric behavior with a sharp field-induced sign reversal of the isothermal magnetic entropy change ($-\Delta S_M$). It exhibits both inverse ($-\Delta S_M < 0$) and conventional ($-\Delta S_M > 0$) magnetocaloric effects. In contrast, MnBi$_4$Te$_7$ shows only conventional magnetocaloric response with a broad positive entropy peak. These results establish structural layering as a key parameter governing magnetic critical fluctuations and magnetocaloric behavior in MnBi$_{2n}$Te$_{3n+1}$ topological magnets.

\end{abstract}

\maketitle


\section{Introduction}
The MnBi$_{2n}$Te$_{3n+1}$ (MBT) family represents a unique class of intrinsic magnetic topological insulators (MTIs) that provide a rich platform to investigate the interplay between magnetism and nontrivial band topology within a van der Waals framework \cite{HasanTI, MBTfamily, MTI, intrinsicMTI, afmTI}. Within this family, MnBi$_2$Te$_4$ ($n=1$) and MnBi$_4$Te$_7$ ($n=2$) are key members \cite{MBTfamily}. MnBi$_2$Te$_4$ crystallizes in a rhombohedral layered structure with the septuple layer (SL) Te--Bi--Te--Mn--Te--Bi--Te as its fundamental building block \cite{Xu2022, Ding2020, STMelectronic}. In this configuration, Mn ions generate strong intralayer ferromagnetic (FM) coupling, while adjacent septuple layers couple antiferromagnetically (AFM), establishing an A-type AFM ground state \cite{STMdistinct, Cui2023, Ge2022, Kim2024}. In contrast, MnBi$_4$Te$_7$ forms a natural van der Waals heterostructure of magnetic MnBi$_2$Te$_4$ septuple and nonmagnetic Bi$_2$Te$_3$ quintuple layer (QL) layers \cite{MBTfamily, recentprogress}. The presence of these nonmagnetic spacer layers weakens the interlayer exchange interactions relative to MnBi$_2$Te$_4$ \cite{guo2024interlayer, Shi2019}. The tunable magnetic states in these layered systems enable diverse topological quantum phenomena including the quantum anomalous Hall effect, axion states and Weyl semimetal phases \cite{Deng2020, axion, Gibertini2019Magnetic2D, layerhalleffect}. Thus, highlighting the MBT family as a central platform for studying topological magnetism \cite{Zhao2021, layertronics}.

Although previous studies have mapped complex magnetic phase diagrams and metamagnetic transitions in detail \cite{zhang2024metamagnetic, das2025metamagnetic, firdosh2025exchange, criticalMnSb}, the precise influence of intercalated non-magnetic quintuple layers and structural complexity on the magnetic critical fluctuations across the MnBi$_{2n}$Te$_{3n+1}$ family remains insufficiently explored \cite{NeutronMBT}. While such structural modulation is known to modify magnetic anisotropy, spin-reorientation behavior, and phase stability \cite{Sun2023, He2022, Klimovskikh2020TunableMagnetism}, its quantitative impact on the divergence of correlation length, the strength of critical spin fluctuations, and the effective magnetic dimensionality near $T_c$ remains unresolved \cite{xuedgeobservation, vidal2019topological}. Resolving this issue is essential because critical fluctuations govern the universality class, entropy redistribution, and the stability of long-range magnetic coherence \cite{nonmagneticSpacer, SpacerMBT}, which in turn directly influences the stability of magnetism-driven topological phenomena such as the QAHE and axion insulators in these layered van der Waals magnetic systems \cite{SpacerMBT, nonmagneticSpacer, Yang2023, Zhao2023}.

\begin{figure*}[t]
\centering
\includegraphics[width=1.01\linewidth]{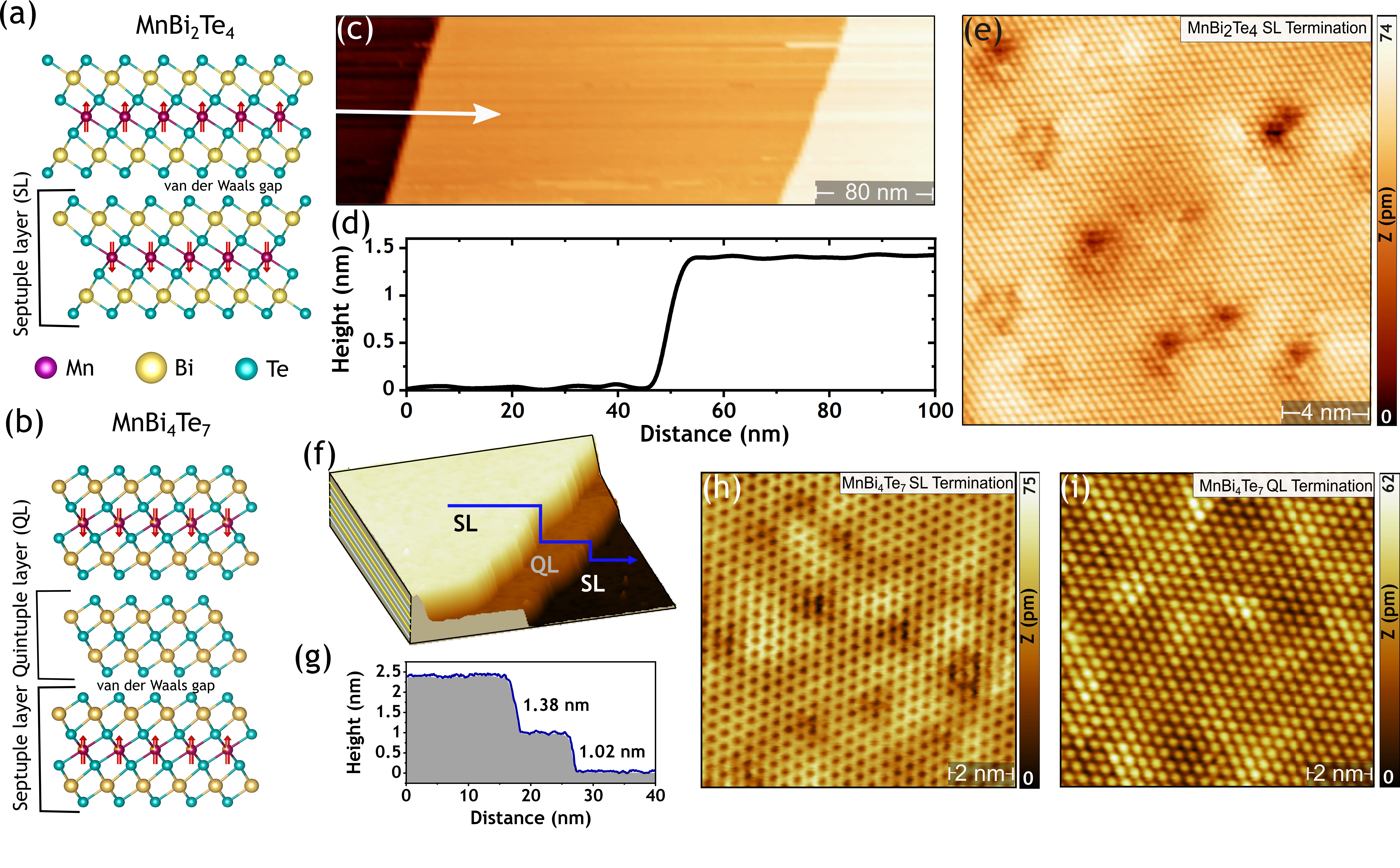}
\caption{(a) Crystal structure of MnBi$_2$Te$_4$ composed of septuple layers (SL) stacked along the $c$-axis.
(b) Crystal structure of MnBi$_4$Te$_7$, where MnBi$_2$Te$_4$ septuple layers alternate with Bi$_2$Te$_3$ quintuple layers (QL). 
(c) Large-scale (400~nm $\times$ 400~nm) constant current STM topography(U = 0.5 V, I = 60 pA) obtained on freshly cleaved (0001) surface of MnBi$_2$Te$_4$, displaying atomically flat terraces separated by sharp step edges. The height profile along the indicated direction yields a step height $\sim$1.41~nm consistent with a single septuple-layer thickness as shown in (d). 
(e) Atomically-resolved STM image (20~nm $\times$ 20~nm) of the SL-terminated surface of MnBi$_2$Te$_4$, clearly depicting the hexagonal lattice of Te-terminated surface (U = 150 mV, I = 300 pA).
(f) Large-scale STM topography (50~nm $\times$ 50~nm) taken (U = 900 mV, I = 140 pA) of the Te-terminated (0001) surface of MnBi$_4$Te$_7$, exhibiting terraces with two discrete step heights corresponding to distinct surface terminations. (g) Line profile across the step in (f), revealing step heights of $\sim$1.38~nm and $\sim$1.02~nm, consistent with septuple- and quintuple-layer thicknesses, respectively. (h) High-resolution STM image (10~nm $\times$ 10~nm) acquired on the SL-terminated terrace (U = 500 mV, I = 460 pA) of MnBi$_4$Te$_7$. (i) Atomic-resolution STM image (10~nm $\times$ 10~nm) acquired (U = 400 mV, I = 150 pA) on the QL-terminated terrace, both depicting the hexagonal Te- terminated surface lattice.}
\end{figure*}

Critical scaling analysis provides a direct and quantitative framework to probe these effects \cite{fisher1967theory, fisher1972critical}. Here, we extract the critical exponents $\beta$, $\gamma$, and $\delta$, which characterize spontaneous magnetization growth, susceptibility divergence and field-dependent scaling near the transition \cite{entropyrcpthetan}. In non-trivial magnetic systems, deviations from ideally known universality classes such as the three-dimensional Ising or Heisenberg values is often reported \cite{chauhan2022different, zhang2016critical, criticalMnSb} These deviations may indicate reduced effective dimensionality, incomplete development of critical fluctuations or competition between antiferromagnetic and field-induced ferromagnetic states \cite{meng2023crossover,VI3critical, fgtcritical1, FGTcritical}. Similarly, in higher-$n$ compounds such as MnBi$_4$Te$_7$, it is still unclear whether the structural dilution in MBT family preserves asymptotic three-dimensional criticality or instead drives a dimensional crossover toward suppressed fluctuations \cite{Liu2022, Liang2023}.

\begin{figure*}[t]
  \centering
  \includegraphics[width=1\linewidth]{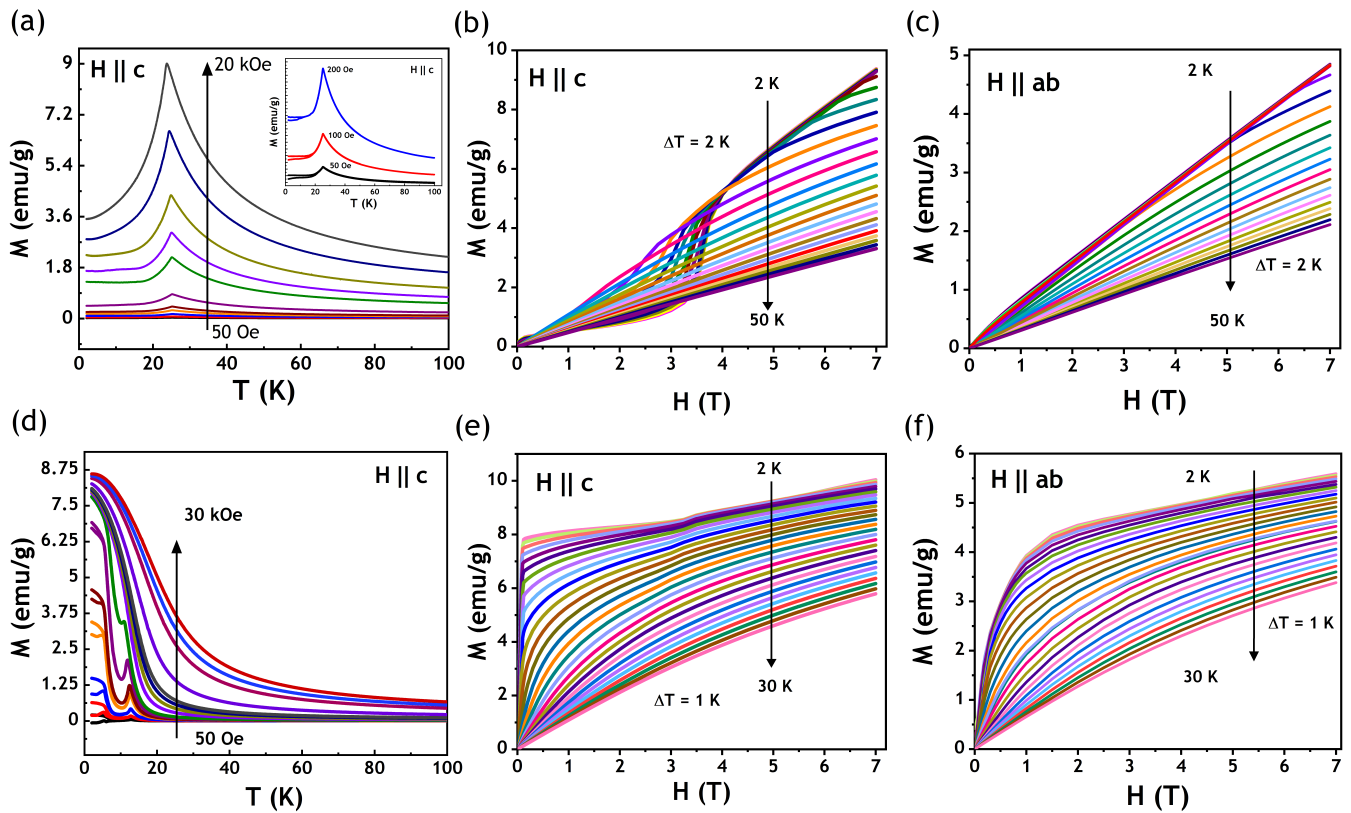}
  \caption{(a) M–T curves of MnBi$_2$Te$_4$ single crystals under different magnetic field with $H \parallel c$. The anomaly at $\sim$24.1 K marks the AFM transition. Inset: zoomed M-T curves for lower fields. (b, c) Isothermal magnetization of MnBi$_2$Te$_4$, for $H \parallel c$ and $H \parallel ab$, respectively. We observe a metamagnetic transition (spin--flip) around 2-4 $T$ in $H \parallel c$, evolving into a gradual response toward $T_N$ in $H \parallel c$ configuration. (d) M–T curves of MnBi$_4$Te$_7$ single crystals under ZFC and FC with $H \parallel c$. The transition at $\sim$12.9 K indicates layered AFM ordering. (e, f) Isothermal magnetization of MnBi$_4$Te$_7$, for $H \parallel c$ and $H \parallel ab$, respectively. We observe a weak anomaly at lower temperatures and magnetic fields compared with MnBi$_2$Te$_4$.  
}
\end{figure*}

\begin{figure*}[t]
  \centering
  \includegraphics[width=0.7\linewidth]{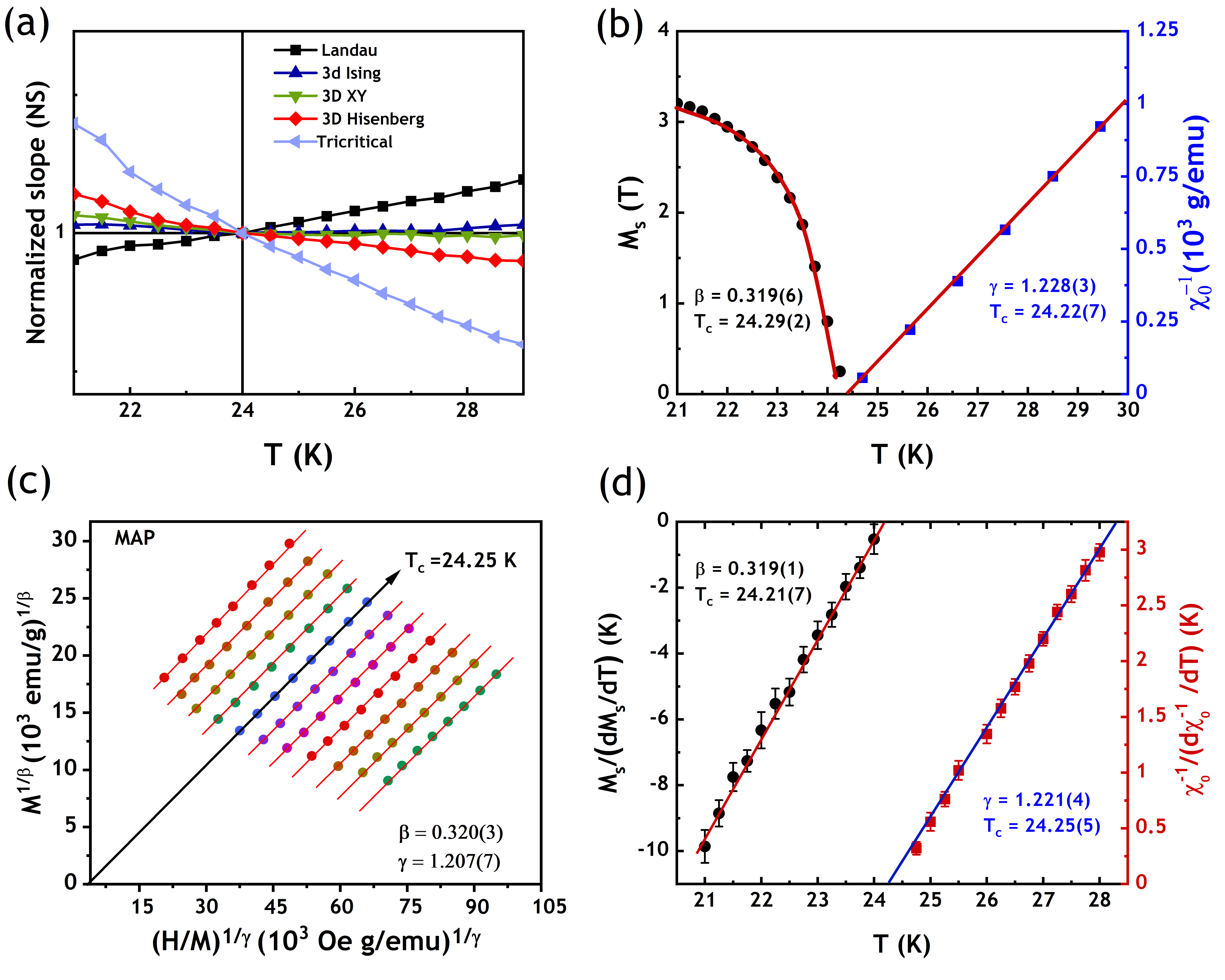}
  \caption{Normalized slope of the Arrott plots as a function of temperature in MnBi$_2$Te$_4$. (b) Temperature dependence of the spontaneous magnetization 
$M_{S}$ (left axis) and the inverse initial susceptibility $\chi_{0}^{-1}$ 
(right axis) fitted with the power-law relations in Eqs.~(3) and (5). 
The solid lines represent the best fits. (c) Modified Arrott plots $(M^{1/\beta}$ vs.\ $(H/M)^{1/\gamma})$ yields $ \beta $=0.319 and $\gamma$=1.221. (d) Kouvel--Fisher plots of $M_{S}(dM_{S}/dT)^{-1}$ 
(left axis) and $\chi_{0}^{-1}(d\chi_{0}^{-1}/dT)^{-1}$ (right axis) as a 
function of temperature.
}
\end{figure*}

Although critical scaling characterizes the evolution of magnetic fluctuations near ordering transitions, it does not directly probe the thermodynamic nature of field-driven instabilities \cite{mcequantative}. The magnetocaloric effect (MCE) provides a complementary thermodynamic probe by quantifying the isothermal entropy change or adiabatic temperature change induced by an applied magnetic field \cite{magnetocaloric}. Therefore, using this complementary technique we can bridge the gap between structural layering, critical scaling behavior, and thermodynamic response in van der Waals MBT family. 

Here, we present a systematic comparative study combining the atomic-resolution scanning tunneling microscopy, magnetic critical behavior and magnetocaloric response in MnBi$_2$Te$_4$ and MnBi$_4$Te$_7$. Our analysis reveals that structural layering, determined by the number of nonmagnetic Bi$_2$Te$_3$ spacer layers, serves as a key parameter controlling the effective magnetic dimensionality and critical universality class. We further show that this structural modulation strongly influences the magnetocaloric response of the two compounds and the nature of the magnetic phase transition across the MnBi$_{2n}$Te$_{3n+1}$ family.

\section{Experimental Methods}

Single crystals of MnBi$_2$Te$_4$ and MnBi$_4$Te$_7$ were grown using self-flux method \cite{firdosh2025exchange, MBTfamily}. High-purity Mn, Bi, and Te elements were weighed in the appropriate stoichiometric ratios, thoroughly mixed, pelletized and sealed in evacuated quartz ampoules. The sealed ampoules were heated to 900~$^\circ$C and slowly cooled to 585~$^\circ$C for MnBi$_2$Te$_4$ and 587~$^\circ$C for MnBi$_4$Te$_7$, followed by rapid quenching in water. The growth yielded shiny, plate-like layered crystals with typical dimensions of approximately $6 \times 4 \times 2$~mm$^3$.
Structural characterization was carried out using powder X-ray diffraction (XRD) and energy-dispersive X-ray (EDX) spectroscopy, confirming phase purity and stoichiometric composition. Magnetic measurements were performed using a Quantum Design MPMS3 magnetometer. Magnetization was measured in both zero-field-cooled (ZFC) and field-cooled (FC) protocols over a temperature range of 2--300~K under applied magnetic fields of up to 7~T. 

To investigate the large-scale surface topography and atomic-scale morphology of the crystals, scanning tunneling microscopy (STM) measurements were performed. All STM experiments were carried out in an ultrahigh vacuum (UHV) commercial Unisoku low-temperature STM system operating at a base pressure of approximately $2 \times 10^{-10}$ mbar \cite{firdosh2025exchange}. Single crystals were cleaved \textit{in situ} to expose fresh (0001) surfaces and were immediately transferred to the STM stage to minimize surface degradation. Measurements were conducted at 78 K using mechanically cut PtIr tips. Topographic images were acquired in the constant-current mode using tunneling set-point currents in the range of 50--500 pA and sample bias voltages between 0.1 and 1.0 V.

\section{Results and discussion}
\subsection{STM imaging of SL–QL layered structures}

To probe the structural evolution across the MnBi$_{2n}$Te$_{3n+1}$ series, scanning tunneling microscopy (STM) measurements were performed on MnBi$_2$Te$_4$ ($n=1$) and MnBi$_4$Te$_7$ ($n=2$) under identical conditions. Large-area constant-current topographs of MnBi$_2$Te$_4$ [Fig.~1(c)] reveal atomically flat terraces separated by step edges of $\sim$1.41~nm. This step height is consistent with the crystallographic septuple-layer (SL) thickness along the $c$ axis, as verified by the line-profile analysis shown in Fig.~1(d). Atomic-resolution topography [Fig.~1(e)] resolves the hexagonal Te surface lattice on the (0001) termination. Thus, confirming the expected SL termination in MnBi$_2$Te$_4$.

  The large-area topography in MnBi$_4$Te$_7$ [Fig.~1(f)] reveals terraces separated by two distinct step heights. The corresponding height profile is shown in Fig.~1(g). Two distinct step heights of approximately 1.38~nm and 1.02~nm are observed. Atomically-resolved STM images acquired on different terraces further confirm the surface terminations. Fig.~1(h) shows an atomic-scale topography obtained on the SL termination, where a well-ordered hexagonal Te surface lattice is clearly resolved. Moreover, Fig.~1(i) shows the atomic-scale topography of the Te- terminated QL surface. Therefore, the atomic-resolution of the SL-terminated Te surface in MnBi$_2$Te$_4$ and coexistence of the SL-QL terminations in MnBi$_4$Te$_7$, provide direct real-space evidence of structural stacking for both compounds and are in good agreement with previous reports \cite{STMelectronic, STMdistinct, firdosh2025exchange}.

\subsection{Critical scaling methodology}
To investigate the nature of the magnetic phase transition in the MBT compounds, we first employed the conventional Arrott plot analysis, where $M^2$ is plotted as a function of $H/M$ based on the mean-field framework. Within this model \cite{arrott1957criterion}, the Arrott plots are expected to display a set of parallel straight lines in the high-field region, with the isotherm at $T_c$ intersecting the origin. The sign of the slope provides a criterion for the transition order: a positive slope corresponds to a continuous (second-order) transition, while a negative slope signals a discontinuous (first-order) transition \cite{banerjee1964generalised}. As shown in Fig.S2(a) and Fig.S3(a), the isotherms for MnBi$_2$Te$_4$ and MnBi$_4$Te$_7$ exhibit positive slopes at high fields, suggesting a second-order magnetic transition. However, the nonlinearity and lack of parallelism in the curves indicate that the mean-field approximation fails to capture the critical behavior of the system.  

\begin{figure*}[t]
  \centering
  \includegraphics[width=1\linewidth]{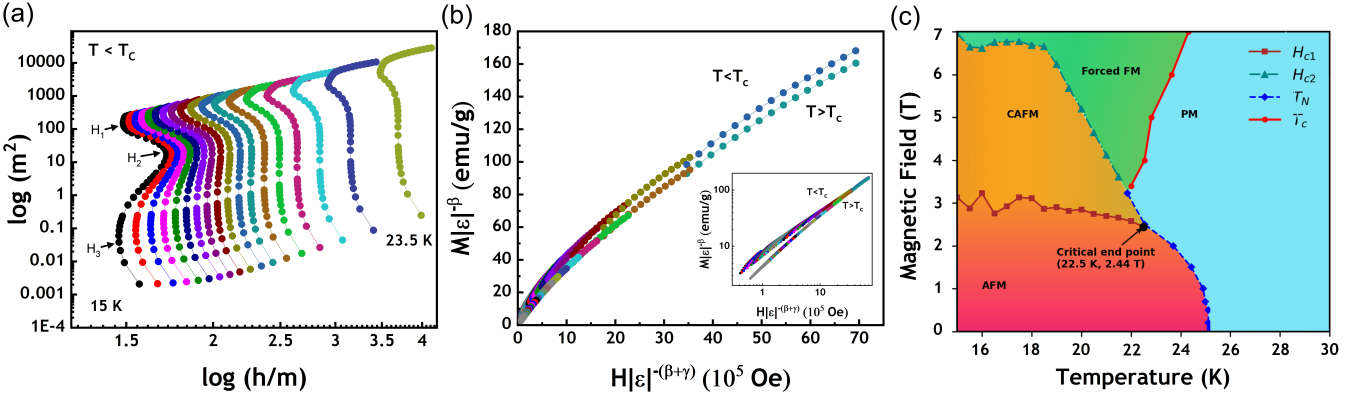}
  \caption{(a) Scaling plots of the renormalized magnetization $M|\varepsilon|^{-\beta}$ versus renormalized field $H|\varepsilon|^{-(\beta+\gamma)}$ for MnBi$_2$Te$_4$, confirms the validity of the critical exponents $\beta = 0.319$ and $\gamma = 1.221$.  The inset shows the same data on a log--log scale, confirming consistency with the scaling hypothesis. 
(b) Enlarged view of the low-temperature region of the log–log Renormalized Arrott Plots, where arrows indicate characteristic phase-boundary points.
(c) Magnetic phase diagram of MnBi$_2$Te$_4$ obtained by combining the results from the renormalized Arrott plot (RAP) analysis with field-dependent magnetization measurements.
}
\end{figure*}

To properly describe the critical region, we applied the Arrott–Noakes equation of state \cite{arrott1967approximate},
\begin{equation}
\left(\frac{H}{M}\right)^{1/\gamma} = a\varepsilon + bM^{1/\beta}, 
\end{equation}
where $\beta$ and $\gamma$ are the critical exponents associated with the spontaneous magnetization and the inverse susceptibility, respectively and $a$ and $b$ are constants. The reduced temperature is defined as
\begin{equation}
\varepsilon = \frac{T}{T_c} - 1
\end{equation}
From scaling theory, the critical behavior is further characterized by the relations
\begin{equation}
M_s(T) \propto (-\varepsilon)^\beta, \quad \varepsilon < 0 \;\; (T<T_c),
\end{equation}
\begin{equation}
M(H) \propto H^{1/\delta}, \quad \varepsilon = 0 \;\; (T=T_c),
\end{equation}
\begin{equation}
\chi_0^{-1}(T) \propto \varepsilon^\gamma, \quad \varepsilon > 0 \;\; (T>T_c)
\end{equation}
where $\beta$, $\gamma$, and $\delta$ correspond to the exponents governing the spontaneous magnetization, the initial susceptibility, and the critical isotherm, respectively.

Since the standard Arrott plots are inadequate, we constructed modified Arrott plots (MAPs) of $M^{1/\beta}$ versus $(H/M)^{1/\gamma}$ using trial values of the exponents corresponding to various universality classes, such as the 2D Ising ($\beta=0.125$, $\gamma=1.75$), 3D Heisenberg ($\beta=0.365$, $\gamma=1.386$), 3D XY ($\beta=0.345$, $\gamma=1.316$), 3D Ising ($\beta=0.325$, $\gamma=1.24$), and tricritical mean-field ($\beta=0.25$, $\gamma=1.0$) (Fig S2 (b - f) and S3 (b - f). To quantify the accuracy of each model, we used the normalized slope (NS) method: $NS(T) = \frac{S(T)}{S(T_C)}, \quad S(T) = \left( \frac{dM^{1/\beta}}{d(H/M)^{1/\gamma}} \right)$. An ideal model would produce parallel isotherms with NS $\approx 1$ near $T_C$. The representative figures for both the samples are shown in Fig. 3(a) and Fig. 5(a), respectively.

The inadequacy of predefined universality classes motivates a model-independent determination of critical exponents. We therefore implemented a modified iterative method (MIM), following approaches developed for systems exhibiting crossover behavior \cite{pramanik2009critical, chauhan2022different}. Starting with trial $\beta$ and $\gamma$ values, we obtained $M_s(T)$ and $\chi_0^{-1}(T)$ by linear extrapolation of the high-field portions of the MAPs to the respective axes. The resulting data were fitted using Eqs. (3) and (5) to refine the exponents. The ordered- and paramagnetic-branch data were then fitted separately to their respective power laws to update $\beta$, $\gamma$, and $T_\mathrm{C}$. The procedure was iterated until convergence, which is defined by stable exponent values and parallel high-field isotherms. This two-sided scheme enables the detection of asymmetric effective exponents and is particularly suited for systems exhibiting crossover phenomena. The modified plots were then reconstructed with the updated values, and the process was repeated until convergence was achieved. This self-consistent method yielded stable values of $\beta$ and $\gamma$, independent of the initial guess, confirming the robustness of the obtained exponents. The final MAPs, shown in Fig.3(b, c) and Fig.5(b), display a series of nearly parallel straight lines in the critical region, thereby validating the reliability of the extracted critical parameters for MnBi$_2$Te$_4$ and MnBi$_4$Te$_7$.

\begin{figure*}[t]
\centering
\includegraphics[width=0.8\linewidth]{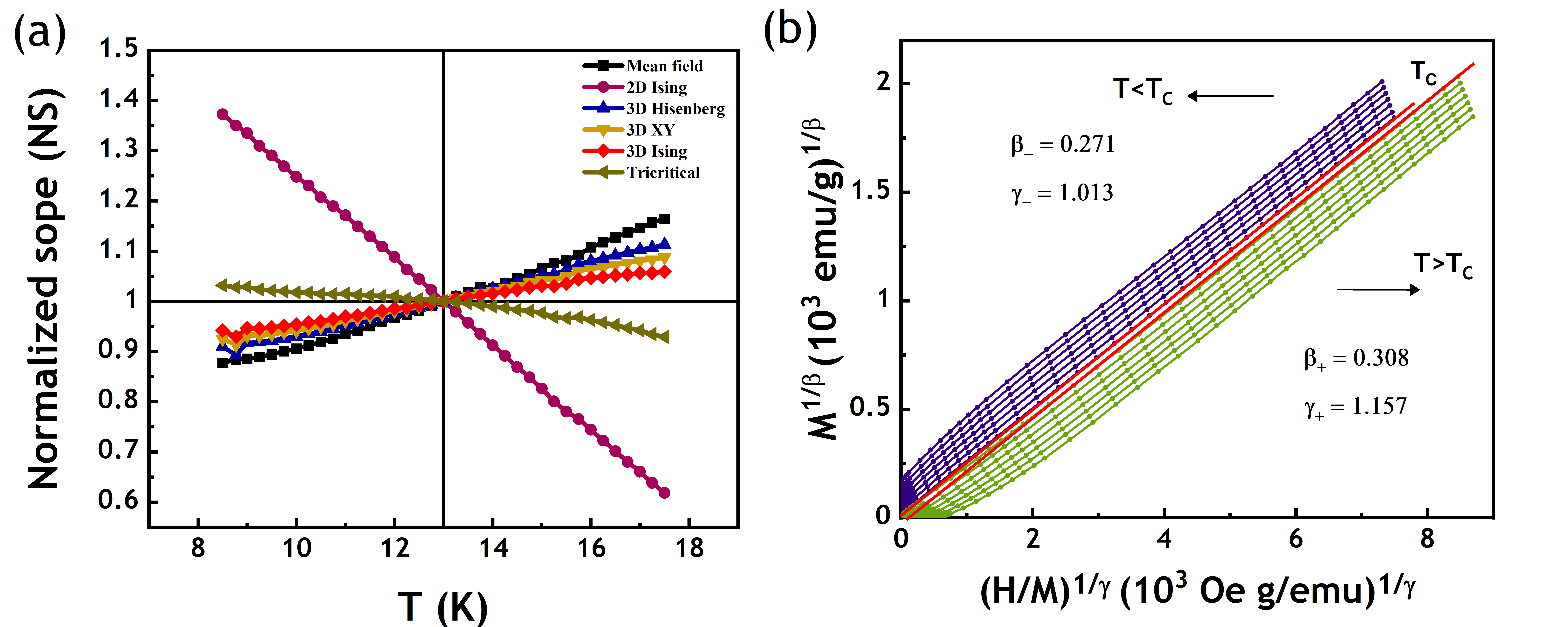}
\caption{(a) Normalized slope of the Arrott plots as a function of temperature in MnBi$_4$Te$_7$.
(b) Modified Arrott plots (MAPs) constructed using the critical exponents obtained from the modified iteration method.}
\end{figure*}

To further validate the exponents, we performed Kouvel–Fisher (KF)
analysis, where the relations  \cite{fisher1972critical}
\begin{equation}
\frac{M_\mathrm{S}(T)}{dM_\mathrm{S}(T)/dT} = \frac{T-T_\mathrm{C}}{\beta},
\end{equation}
\begin{equation}
\frac{\chi_0^{-1}(T)}{d\chi_0^{-1}(T)/dT} = \frac{T-T_\mathrm{C}}{\gamma},
\end{equation}

yield straight lines whose slopes provide independent estimates of
$\beta$ and $\gamma$. The KF plots [Fig.3(d) and S4(c)] are in good agreement with the MAP results. The critical isotherm at $T_\mathrm{C}$, plotted on a log--log scale follows the relation

\begin{equation}
M \propto H^{1/\delta},
\end{equation}

from which we obtain $\delta$. The consistency of the
Critical exponents can be further checked using the Widom relation,

\begin{equation}
\delta = 1 + \frac{\gamma}{\beta},
\end{equation}

We verified the reliability of the critical exponents by performing a scaling analysis of the magnetization isotherms near $T_{c}$. According to the scaling hypothesis, the equation of state close to the critical point can be written as  
\begin{equation}
M(H,\varepsilon) = |\varepsilon|^{\beta} f_{\pm}\!\left(\frac{H}{|\varepsilon|^{\beta+\gamma}}\right),
\end{equation}  
where $\varepsilon = (T - T_{c})/T_{c}$ is the reduced temperature and $f_{\pm}$ denotes the scaling functions for $T<T_{c}$ and $T>T_{c}$. In Fig.~4(a) and Fig. 6(a), we plot $M|\varepsilon|^{-\beta}$ as a function of $H|\varepsilon|^{-(\beta+\gamma)}$ for MnBi$_{2}$Te$_{4}$ and MnBi$_{4}$Te$_{7}$, respectively. While Fig.~6(b) shows the renormalized form $m^{2}$ versus $h/m$, where $m = M|\varepsilon|^{-\beta}$ and $h = H|\varepsilon|^{-(\beta+\gamma)}$. The collapse of the isotherms into two distinct branches corresponding to $T<T_{c}$ and $T>T_{c}$ confirms the internal consistency of the extracted critical exponents. This excellent scaling behavior provides compelling evidence that the magnetic phase transition in both samples around the transition temperature is of second order, in agreement with the predictions of the scaling theory.

As we know, for a homogeneous magnet, the universality class of the magnetic
phase transition depends on the nature of the exchange interaction $J(r)$. 
Fisher \textit{et al.} theoretically treated this kind of ordering as an attractive 
interaction of spins, where a renormalization-group analysis suggests that the exchange interaction decays with distance $r$ as \cite{fisher1972critical}
\begin{equation}
J(r) \sim r^{-(d+\sigma)},
\end{equation}
where $\sigma$ is a positive constant. Moreover, the susceptibility exponent 
$\gamma$ is predicted as \cite{fisher1972critical}.
\begin{equation}
\begin{split}
\gamma = 1 &+ \tfrac{4}{d}\!\left(\tfrac{n+2}{\,n+8\,}\right)\Delta\sigma \\
&+ \tfrac{8(n+2)(n-4)}{d^{2}(n+8)^{2}}
\!\left[1+\tfrac{2G(d/2)}{d}\right]
\tfrac{(7n+20)}{(n-4)(n+8)}(\Delta\sigma)^{2}.
\end{split}
\end{equation}
where $\Delta\sigma = \sigma - d/2$, $G(d/2) = 3 - \tfrac{1}{4}(d/2)^2$, 
$d$ is the spatial dimension, and $n$ is the spin dimensionality.

\subsection{Critical behavior of MnBi$_2$Te$_4$ and MnBi$_4$Te$_7$}

\begin{figure*}[t]
\centering
\includegraphics[width=0.8\linewidth]{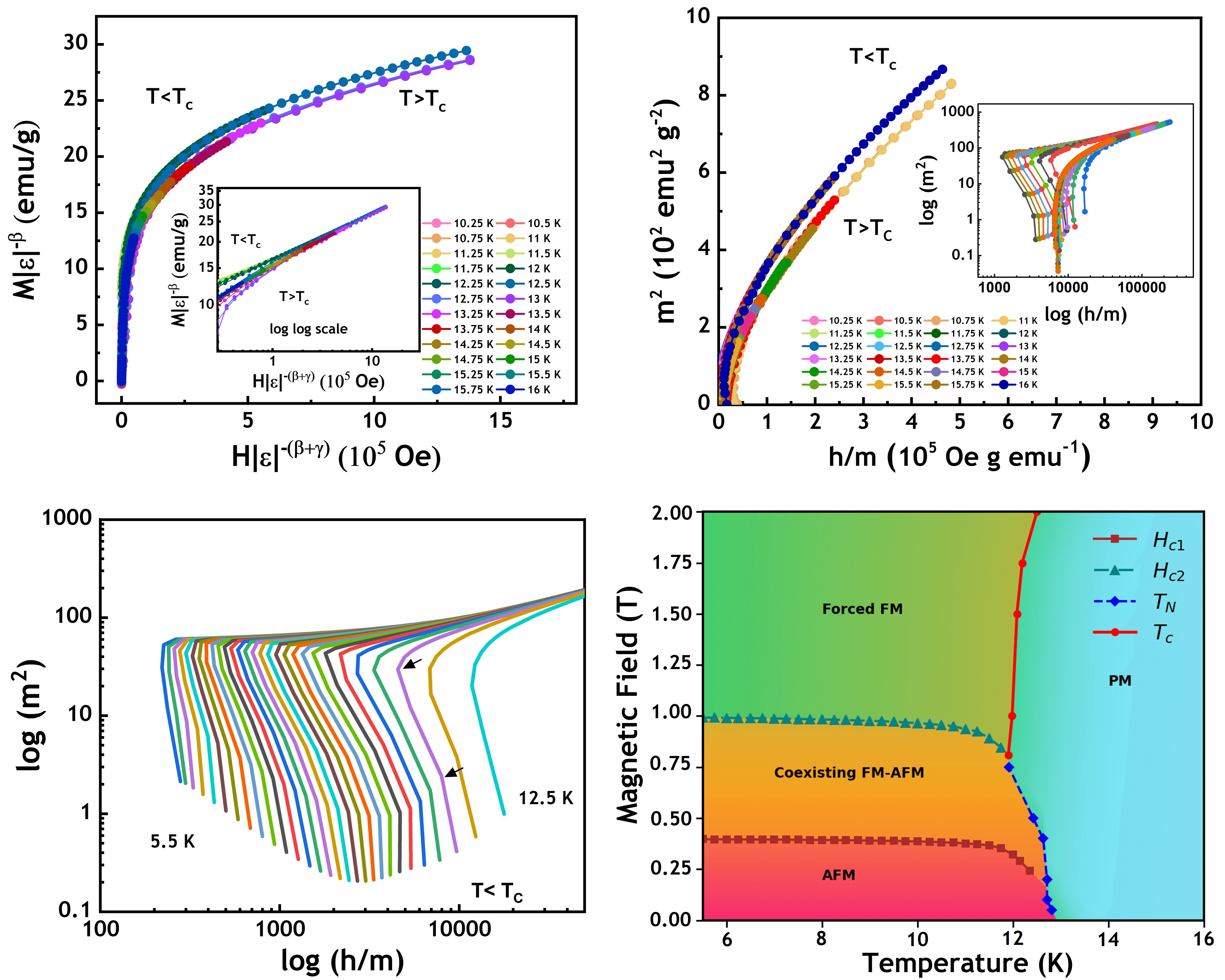}
\caption{(a) Scaled magnetization data of MnBi$_4$Te$_7$ analyzed within the framework of the magnetic equation of state. The data collapse onto two separate branches below and above $T_c$, confirms the validity of the critical exponents $\beta^- \approx 0.271$, $\gamma^- \approx 1.013$ below $T_c$ and $\beta^+ \approx 0.308$, $\gamma^+ \approx 1.157$ above $T_c$.  
The inset shows the same data on a log--log scale, which confirms consistency with the scaling hypothesis. 
(b) Scaling collapse of the renormalized magnetization and field, $m^2$ vs $h^2$, is in good agreement for both $T<T_c$ and $T>T_c$.
(c) Enlarged view of the low-temperature region of the log–log RAP, where arrows schematically indicate characteristic phase-boundary points.
(d) Magnetic phase diagram of MnBi$_4$Te$_7$.
}
\end{figure*}

In MnBi$_2$Te$_4$, the spontaneous magnetization $M_s(T)$ and inverse initial susceptibility $\chi_0^{-1}(T)$ were obtained from the intercepts of the high-field Arrott isotherms [Fig.~3(b)] and analyzed using the Kouvel--Fisher method [Fig.~3(d)]. This analysis yields the critical exponents $\beta \approx 0.319$ and $\gamma \approx 1.221$, together with a consistent transition temperature $T_c \approx 24.25$~K. The reliability of these exponents is further supported by scaling analysis, where the magnetization data collapse onto two universal branches above and below $T_c$ as shown in Fig.~4(a). The obtained values are close to those expected for the three-dimensional Ising universality class, indicating strong easy-axis anisotropy in MnBi$_2$Te$_4$. Further insight into the field evolution of the magnetic state is obtained from the renormalized Arrott plots shown on a log–log scale in Fig.~4(b). The systematic evolution of the curves with temperature reflects the field-driven modification of the antiferromagnetic order. Combining these results with field-dependent magnetization measurements allows the construction of the magnetic phase diagram of MnBi$_2$Te$_4$, Fig.~4(c).

\begin{figure*}[t]
\centering
\includegraphics[width=0.9\linewidth]{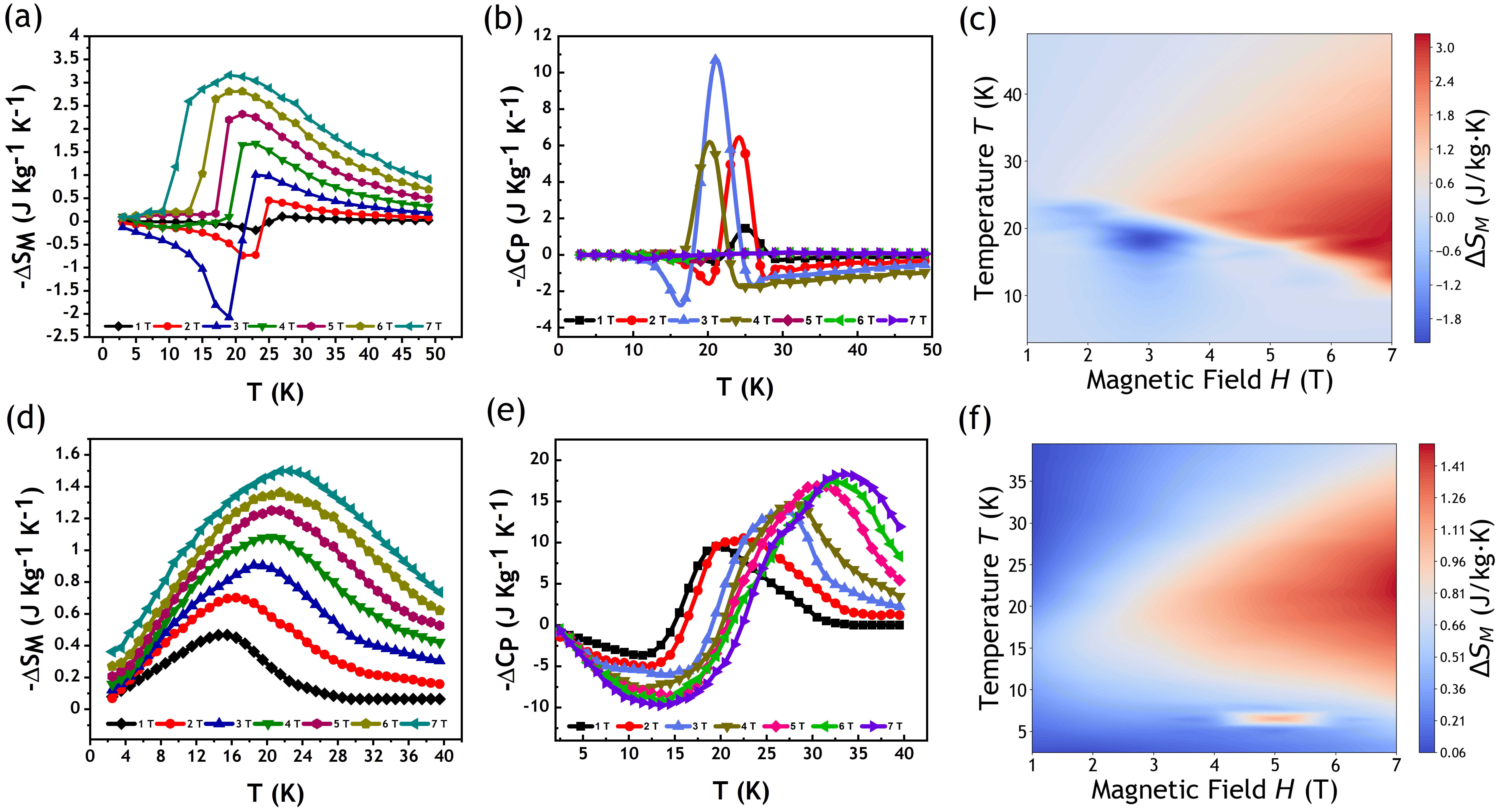}
\caption{
(a, d) Temperature dependence of the magnetic entropy change $-\Delta S_M(T, H)$ for MnBi$_2$Te$_4$ and MnBi$_4$Te$_7$, respectively, calculated from magnetization isotherms for fields ranging from 1 to 7 T. MnBi$_2$Te$_4$ exhibits a dual magnetocaloric response with a positive entropy peak near $T_c (\approx 24.25$ K) and a inverse entropy pocket at lower temperatures associated with field-induced spin-flip transitions. In contrast, MnBi$_4$Te$_7$ shows a broad and predominantly positive entropy change centered near $T_c (\approx 13.05$ K). 
(b, e) Temperature dependence of the field-induced heat-capacity change $-\Delta C_P(T, H)$ for MnBi$_2$Te$_4$ and MnBi$_4$Te$_7$, respectively. MnBi$_2$Te$_4$ displays sharp field-enhanced anomalies near $T_c$, whereas MnBi$_4$Te$_7$ exhibits broader variations extending over a wider temperature range. 
(c, f) Contour maps of the magnetic entropy change $\Delta S_M(T, H)$ for MnBi$_2$Te$_4$ and MnBi$_4$Te$_7$, illustrating the evolution of entropy with temperature and magnetic field. The contour map of MnBi$_2$Te$_4$ reveals a distinct region of negative entropy change at intermediate fields and low temperatures, while MnBi$_4$Te$_7$ shows a smooth, predominantly positive entropy distribution.
}
\end{figure*}
The critical behavior of MnBi$_4$Te$_7$ differs from that of MnBi$_2$Te$_4$, reflecting the reduced magnetic connectivity introduced by the insertion of nonmagnetic Bi$_2$Te$_3$ quintuple layers. Separate fits above and below $T_c$ yield distinct effective critical exponents: for $T<T_c$, $\beta^- \approx 0.271$ and $\gamma^- \approx 1.013$, while for $T>T_c$ we obtain $\beta^+ \approx 0.308$ and $\gamma^+ \approx 1.157$ Fig.~5(b). The asymmetry of the exponent sets indicates that the transition is not governed by a single universal class.

Furthermore, Fig.~6 depicts the scaling analysis of the magnetization data, which supports this conclusion. A single set of critical exponents fails to collapse the data over the entire temperature range [Fig.~S4(a)], whereas separate scaling above and below $T_c$ yields improved agreement [Fig.~5(b)]. Such behavior is characteristic of crossover-dominated criticality, which can be attributed to weakened interlayer exchange and the proximity of a low-field tricritical point reported near $11.9$~K and $750$~Oe \cite{zhang2024metamagnetic}. The combined influence of reduced magnetic coherence and tricritical fluctuations suppresses the asymptotic critical region, resulting in effective exponent values that vary across the transition.

Moreover, our previous work \cite{firdosh2025exchange} reveals a frequency-dependent anomaly near 6~K, indicating freezing of ferromagnetic clusters within an antiferromagnetic matrix. Thus, while the transition near $T_c$ reflects antiferromagnetic ordering, the low-temperature state is governed by cluster-driven magnetic dynamics. 

To probe the spatial range of magnetic interactions, we analyze the critical exponents within the renormalization-group framework for long-range exchange $J(r)\sim r^{-(d+\sigma)}$. Assuming $d=3$ and Ising spin symmetry ($n=1$), the measured exponents yield $\sigma \approx 1.88$ for MnBi$_2$Te$_4$, corresponding to $J(r)\sim r^{-4.88}$. For MnBi$_4$Te$_7$, the values $\beta_- \approx 0.271$, $\gamma_- \approx 1.013$ give $\sigma_- \approx 1.53$ below $T_c$, while $\beta_+ \approx 0.308$, $\gamma_+ \approx 1.157$ yield $\sigma_+ \approx 1.80$ above $T_c$, corresponding to $J(r)\sim r^{-4.53}$ and $J(r)\sim r^{-4.80}$, respectively.

The range $1.5<\sigma<2$ places both systems in the crossover regime between long-range and short-range magnetic interactions, indicating intermediate-range exchange. Moreover, we observe a slower spatial decay of the exchange interaction in MnBi$_4$Te$_7$  compared with MnBi$_2$Te$_4$, implying a slightly longer effective interaction range in MnBi$_4$Te$_7$. This behavior likely originates from the insertion of the Bi$_2$Te$_3$ quintuple layer, which modifies the interlayer exchange pathways within the MnBi$_{2n}$Te$_{3n+1}$ structure.

\subsection{Magnetocaloric signatures of MnBi$_2$Te$_4$ and MnBi$_4$Te$_7$}

The magnetocaloric effect (MCE) originates from the coupling between the magnetic entropy and the external magnetic field, and provides valuable information about the nature of magnetic interactions and phase transitions. The isothermal magnetic entropy change, $\Delta S_M(T, H)$, can be obtained from the magnetization isotherms using Maxwell’s relation, which connects the entropy S with the magnetization M as \cite{entropy, entropy2}
\begin{equation}
\left(\frac{\partial S}{\partial H}\right)_T = \left(\frac{\partial M}{\partial T}\right)_H.
\end{equation}
Integrating over the magnetic field yields
\begin{equation}
\Delta S_M(T, \Delta H) = \int_{0}^{\Delta H} \left( \frac{\partial M}{\partial T} \right)_H dH,
\end{equation}
where $\Delta H$ is the change in magnetic field. In practice, for a discrete set of isothermal $M(H)$ curves measured at successive temperatures, the entropy change $-\Delta S_{M}(T,H)$  can be approximated as \cite{entropy2}
\begin{equation}
\left|\Delta S_{M}\left(\frac{T_{i}+T_{i+1}}{2},H\right)\right| = \sum_{i} \left[\frac{(M_{i+1}-M_{i})}{T_{i+1}-T_{i}}\right]\Delta H_{i}.
\end{equation}

Also, the magnetic contribution to the specific heat can be derived from entropy as \cite{entropy2}
\begin{equation}
\Delta C_{p}(T,H) = C_{p}(T,H) - C_{p}(T,0) = T \frac{\partial \Delta S_{M}(T,H)}{\partial T}.
\end{equation}

For MnBi$_2$Te$_4$ [Fig.~7(a)], the isothermal magnetic entropy change $-\Delta S_M(T)$ displays a dual behaviour. At low applied fields ($\mu_0H = 1$--$3$~T), we observe a negative entropy contribution around 15-18 K, indicating an inverse magnetocaloric effect. As the field is increased ($\mu_0H = 4$--$7$~T), the entropy change undergoes a well-defined sign inversion and shows conventional magnetocaloric effect with a sharp positive peak around $T_c \sim 24.25$. For a maximum field change $\Delta H = 7$~T the peak magnitude reaches $-\Delta S_M^{\mathrm{max}} \approx 3$~J\,kg$^{-1}$K$^{-1}$. The pronounced asymmetry and field-dependent switching of entropy change is consistent with a field-induced reorientation spin-flip transition of MnBi$_2$Te$_4$, which produces a rapid redistribution of magnetic entropy. The contour map of $-\Delta S_M(T, H)$ [Fig.~7(c)] further highlights this behavior by revealing a distinct region of negative entropy change within the same temperature and field window. 

In contrast, MnBi$_4$Te$_7$ [Fig.~7(d)] exhibits a qualitatively different magnetocaloric behavior. The entropy change remains positive across the measured temperature range and evolves smoothly with increasing field, producing a single, broad maximum centered near the $T_c \sim 13.05$~K. The peak magnitude increases monotonically with field and attains $-\Delta S_M^{\mathrm{max}} \approx 1.5$~J\,kg$^{-1}$K$^{-1}$ for $\Delta H = 7$~T.  The absence of entropy sign reversal and the broad temperature distribution of the entropy peak indicate a more gradual field-driven polarization of moments rather than an abrupt metamagnetic reorientation. The corresponding contour map [Fig.~7(f)] indicates that the entropy change remains predominantly positive throughout the explored temperature and field range. The absence of negative entropy regions and sharp thermodynamic anomalies suggests that field-induced spin-flip processes are significantly suppressed in MnBi$_4$Te$_7$.

These entropy trends are further supported by the field-dependent heat-capacity change $\Delta C_P(T, H)$. In MnBi$_2$Te$_4$ [Fig.~7(b)], $\Delta C_P$ shows sharp field-enhanced peaks near the magnetic ordering temperature ($\sim$20–25 K). This behavior reflects strong thermodynamic fluctuations close to the transition. In contrast, MnBi$_4$Te$_7$ [Fig.~7(e)] shows a broader and more gradual variation of $\Delta C_P(T,H)$ over a wider temperature range ($\sim$10–35 K). The heat-capacity change increases smoothly with field and does not show the sharp anomalies seen in MnBi$_2$Te$_4$. This broader response indicates that magnetic entropy is redistributed over a wider temperature range. Such behavior is consistent with reduced effective interlayer coupling and the coexistence of ferromagnetic and antiferromagnetic correlations in MnBi$_4$Te$_7$ reported in earlier studies \cite{firdosh2025exchange}.

From an application perspective, the sharp entropy variations of MnBi$_2$Te$_4$ make it a strong candidate for low-temperature magnetocaloric cooling and field-tunable spintronic switches that exploit discrete transitions. MnBi$_4$Te$_7$, with smoother tunability and reduced coercivity, is better suited for reversible, low-energy manipulation of topological states. Together, these results establish the MBT family as a versatile materials platform where layer engineering directly controls the interplay between magnetism, entropy, and topology, enabling pathways to realize QAHE and related quantum phases in van der Waals magnets.

\section{Conclusion}

We presented the critical behavior and magnetocaloric properties of MnBi$_2$Te$_4$ and MnBi$_4$Te$_7$ to elucidate how structural layering in the MnBi$_{2n}$Te$_{3n+1}$ series controls magnetic interactions. STM reveals atomically flat septuple-layer terraces in MnBi$_2$Te$_4$, whereas coexisting septuple and quintuple layer terminations in MnBi$_4$Te$_7$. Critical-scaling analysis of the magnetization data yields distinct sets of critical exponents for the two compounds, which indicates different effective ranges of magnetic exchange and therefore different universality classes. Furthermore, magnetocaloric measurements provide complementary thermodynamic evidence. MnBi$_2$Te$_4$ displays both conventional and inverse magnetocaloric responses, whereas MnBi$_4$Te$_7$ displays a single, broad maximum and smoothly varying features in the entropy change ($-\Delta S_M(T, H)$). Altogether, the combined critical scaling and magnetocaloric results identify structural layering as a tunable parameter that controls magnetic dimensionality, the effective interaction range, and the thermodynamic pathways for entropy redistribution in this family. Extending the same analysis to higher-$n$ members of MnBi$_{2n}$Te$_{3n+1}$ can show how the family evolves from three-dimensional criticality toward quasi-two-dimensional crossover behavior and will have direct implications for magnetically driven topological phenomena such as QAHE and axion electrodynamics in these materials.

\section{ACKNOWLEDGMENTS}
We thank Subhasis Ghosh (JNU), Martin Weigel and Christoph Tegenkamp (TUC) for insightful discussions. N.F. acknowledges financial support from the DST INSPIRE Fellowship (IF200582), Government of India. N.F and S.M. acknowledge partial support of the Indo-German project SPARC-GIANT. The authors gratefully acknowledge the Department of Physics and the Central Research Facility (CRF), IIT Delhi, for MPMS and STM measurements.

\bibliographystyle{apsrev4-2-with-title}
\bibliography{references}

\begin{thebibliography}{54}%
\makeatletter
\providecommand \@ifxundefined [1]{%
 \@ifx{#1\undefined}
}%
\providecommand \@ifnum [1]{%
 \ifnum #1\expandafter \@firstoftwo
 \else \expandafter \@secondoftwo
 \fi
}%
\providecommand \@ifx [1]{%
 \ifx #1\expandafter \@firstoftwo
 \else \expandafter \@secondoftwo
 \fi
}%
\providecommand \natexlab [1]{#1}%
\providecommand \enquote  [1]{``#1''}%
\providecommand \bibnamefont  [1]{#1}%
\providecommand \bibfnamefont [1]{#1}%
\providecommand \citenamefont [1]{#1}%
\providecommand \href@noop [0]{\@secondoftwo}%
\providecommand \href [0]{\begingroup \@sanitize@url \@href}%
\providecommand \@href[1]{\@@startlink{#1}\@@href}%
\providecommand \@@href[1]{\endgroup#1\@@endlink}%
\providecommand \@sanitize@url [0]{\catcode `\\12\catcode `\$12\catcode `\&12\catcode `\#12\catcode `\^12\catcode `\_12\catcode `\%12\relax}%
\providecommand \@@startlink[1]{}%
\providecommand \@@endlink[0]{}%
\providecommand \url  [0]{\begingroup\@sanitize@url \@url }%
\providecommand \@url [1]{\endgroup\@href {#1}{\urlprefix }}%
\providecommand \urlprefix  [0]{URL }%
\providecommand \Eprint [0]{\href }%
\providecommand \doibase [0]{https://doi.org/}%
\providecommand \selectlanguage [0]{\@gobble}%
\providecommand \bibinfo  [0]{\@secondoftwo}%
\providecommand \bibfield  [0]{\@secondoftwo}%
\providecommand \translation [1]{[#1]}%
\providecommand \BibitemOpen [0]{}%
\providecommand \bibitemStop [0]{}%
\providecommand \bibitemNoStop [0]{.\EOS\space}%
\providecommand \EOS [0]{\spacefactor3000\relax}%
\providecommand \BibitemShut  [1]{\csname bibitem#1\endcsname}%
\let\auto@bib@innerbib\@empty
\bibitem [{\citenamefont {Hasan}\ and\ \citenamefont {Kane}(2010)}]{HasanTI}%
  \BibitemOpen
  \bibfield  {author} {\bibinfo {author} {\bibfnamefont {M.~Z.}\ \bibnamefont {Hasan}}\ and\ \bibinfo {author} {\bibfnamefont {C.~L.}\ \bibnamefont {Kane}},\ }\bibinfo {title} {Colloquium: Topological insulators},\ \href {https://doi.org/10.1103/RevModPhys.82.3045} {\bibfield  {journal} {\bibinfo  {journal} {Rev. Mod. Phys.}\ }\textbf {\bibinfo {volume} {82}},\ \bibinfo {pages} {3045} (\bibinfo {year} {2010})}\BibitemShut {NoStop}%
\bibitem [{\citenamefont {He}(2020)}]{MBTfamily}%
  \BibitemOpen
  \bibfield  {author} {\bibinfo {author} {\bibfnamefont {K.}~\bibnamefont {He}},\ }\bibinfo {title} {{MnBi\textsubscript{2}Te\textsubscript{4}}-family intrinsic magnetic topological materials},\ \href {https://doi.org/10.1038/s41535-020-00291-5} {\bibfield  {journal} {\bibinfo  {journal} {npj Quantum Materials}\ }\textbf {\bibinfo {volume} {5}},\ \bibinfo {pages} {90} (\bibinfo {year} {2020})}\BibitemShut {NoStop}%
\bibitem [{\citenamefont {Tokura}\ \emph {et~al.}(2019)\citenamefont {Tokura}, \citenamefont {Yasuda},\ and\ \citenamefont {Tsukazaki}}]{MTI}%
  \BibitemOpen
  \bibfield  {author} {\bibinfo {author} {\bibfnamefont {Y.}~\bibnamefont {Tokura}}, \bibinfo {author} {\bibfnamefont {K.}~\bibnamefont {Yasuda}},\ and\ \bibinfo {author} {\bibfnamefont {A.}~\bibnamefont {Tsukazaki}},\ }\bibinfo {title} {Magnetic topological insulators},\ \href {https://doi.org/10.1038/s42254-018-0011-5} {\bibfield  {journal} {\bibinfo  {journal} {Nat. Rev. Phys.}\ }\textbf {\bibinfo {volume} {1}},\ \bibinfo {pages} {126} (\bibinfo {year} {2019})}\BibitemShut {NoStop}%
\bibitem [{\citenamefont {Li}\ \emph {et~al.}(2019)\citenamefont {Li}, \citenamefont {Li}, \citenamefont {Duan}, \citenamefont {Xu},\ and\ \citenamefont {Zhang}}]{intrinsicMTI}%
  \BibitemOpen
  \bibfield  {author} {\bibinfo {author} {\bibfnamefont {J.}~\bibnamefont {Li}}, \bibinfo {author} {\bibfnamefont {Y.}~\bibnamefont {Li}}, \bibinfo {author} {\bibfnamefont {W.}~\bibnamefont {Duan}}, \bibinfo {author} {\bibfnamefont {Y.}~\bibnamefont {Xu}},\ and\ \bibinfo {author} {\bibfnamefont {S.-C.}\ \bibnamefont {Zhang}},\ }\bibinfo {title} {Intrinsic magnetic topological insulators in van der waals layered {MnBi\textsubscript{2}Te\textsubscript{4}}-family materials},\ \href {https://doi.org/10.1126/sciadv.aaw5685} {\bibfield  {journal} {\bibinfo  {journal} {Science Advances}\ }\textbf {\bibinfo {volume} {5}},\ \bibinfo {pages} {eaaw5685} (\bibinfo {year} {2019})}\BibitemShut {NoStop}%
\bibitem [{\citenamefont {Otrokov}\ \emph {et~al.}(2019)\citenamefont {Otrokov}, \citenamefont {Klimovskikh}, \citenamefont {Bentmann}, \citenamefont {Estyunin}, \citenamefont {Zeugner}, \citenamefont {Aliev}, \citenamefont {Gaß}, \citenamefont {Wolter}, \citenamefont {Koroleva}, \citenamefont {Shikin}, \citenamefont {Blanco-Rey}, \citenamefont {Hoffmann}, \citenamefont {Rusinov}, \citenamefont {Vergniory}, \citenamefont {Kimura}, \citenamefont {Petaccia}, \citenamefont {Di~Santo}, \citenamefont {Vidal}, \citenamefont {Schatz}, \citenamefont {Kißner}, \citenamefont {Ünzelmann}, \citenamefont {Min}, \citenamefont {Moser}, \citenamefont {Peixoto}, \citenamefont {Reinert}, \citenamefont {Ernst}, \citenamefont {Echenique}, \citenamefont {Isaeva},\ and\ \citenamefont {Chulkov}}]{afmTI}%
  \BibitemOpen
  \bibfield  {author} {\bibinfo {author} {\bibfnamefont {M.~M.}\ \bibnamefont {Otrokov}}, \bibinfo {author} {\bibfnamefont {I.~I.}\ \bibnamefont {Klimovskikh}}, \bibinfo {author} {\bibfnamefont {H.}~\bibnamefont {Bentmann}}, \bibinfo {author} {\bibfnamefont {D.}~\bibnamefont {Estyunin}}, \bibinfo {author} {\bibfnamefont {A.}~\bibnamefont {Zeugner}}, \bibinfo {author} {\bibfnamefont {Z.~S.}\ \bibnamefont {Aliev}}, \bibinfo {author} {\bibfnamefont {S.}~\bibnamefont {Gaß}}, \bibinfo {author} {\bibfnamefont {A.~U.~B.}\ \bibnamefont {Wolter}}, \bibinfo {author} {\bibfnamefont {A.~V.}\ \bibnamefont {Koroleva}}, \bibinfo {author} {\bibfnamefont {A.~M.}\ \bibnamefont {Shikin}}, \bibinfo {author} {\bibfnamefont {M.}~\bibnamefont {Blanco-Rey}}, \bibinfo {author} {\bibfnamefont {M.}~\bibnamefont {Hoffmann}}, \bibinfo {author} {\bibfnamefont {I.~P.}\ \bibnamefont {Rusinov}}, \bibinfo {author} {\bibfnamefont {M.~G.}\ \bibnamefont {Vergniory}}, \bibinfo {author} {\bibfnamefont {A.}~\bibnamefont {Kimura}}, \bibinfo
  {author} {\bibfnamefont {L.}~\bibnamefont {Petaccia}}, \bibinfo {author} {\bibfnamefont {G.}~\bibnamefont {Di~Santo}}, \bibinfo {author} {\bibfnamefont {R.~C.}\ \bibnamefont {Vidal}}, \bibinfo {author} {\bibfnamefont {S.}~\bibnamefont {Schatz}}, \bibinfo {author} {\bibfnamefont {K.}~\bibnamefont {Kißner}}, \bibinfo {author} {\bibfnamefont {M.}~\bibnamefont {Ünzelmann}}, \bibinfo {author} {\bibfnamefont {C.~H.}\ \bibnamefont {Min}}, \bibinfo {author} {\bibfnamefont {S.}~\bibnamefont {Moser}}, \bibinfo {author} {\bibfnamefont {T.~R.~F.}\ \bibnamefont {Peixoto}}, \bibinfo {author} {\bibfnamefont {F.}~\bibnamefont {Reinert}}, \bibinfo {author} {\bibfnamefont {A.}~\bibnamefont {Ernst}}, \bibinfo {author} {\bibfnamefont {P.~M.}\ \bibnamefont {Echenique}}, \bibinfo {author} {\bibfnamefont {A.}~\bibnamefont {Isaeva}},\ and\ \bibinfo {author} {\bibfnamefont {E.~V.}\ \bibnamefont {Chulkov}},\ }\bibinfo {title} {Prediction and observation of an antiferromagnetic topological insulator},\ \href
  {https://doi.org/10.1038/s41586-019-1840-9} {\bibfield  {journal} {\bibinfo  {journal} {Nature}\ }\textbf {\bibinfo {volume} {576}},\ \bibinfo {pages} {416} (\bibinfo {year} {2019})}\BibitemShut {NoStop}%
\bibitem [{\citenamefont {Xu}\ \emph {et~al.}(2022{\natexlab{a}})\citenamefont {Xu}, \citenamefont {Yang}, \citenamefont {Wang}, \citenamefont {Guzman}, \citenamefont {Gao}, \citenamefont {Zhu}, \citenamefont {Peng}, \citenamefont {Zang}, \citenamefont {Xi}, \citenamefont {Tian}, \citenamefont {Li}, \citenamefont {Lei}, \citenamefont {Luo}, \citenamefont {Yang}, \citenamefont {Wang}, \citenamefont {Xia}, \citenamefont {Zhou}, \citenamefont {Huang},\ and\ \citenamefont {Ye}}]{Xu2022}%
  \BibitemOpen
  \bibfield  {author} {\bibinfo {author} {\bibfnamefont {X.}~\bibnamefont {Xu}}, \bibinfo {author} {\bibfnamefont {S.}~\bibnamefont {Yang}}, \bibinfo {author} {\bibfnamefont {H.}~\bibnamefont {Wang}}, \bibinfo {author} {\bibfnamefont {R.}~\bibnamefont {Guzman}}, \bibinfo {author} {\bibfnamefont {Y.}~\bibnamefont {Gao}}, \bibinfo {author} {\bibfnamefont {Y.}~\bibnamefont {Zhu}}, \bibinfo {author} {\bibfnamefont {Y.}~\bibnamefont {Peng}}, \bibinfo {author} {\bibfnamefont {Z.}~\bibnamefont {Zang}}, \bibinfo {author} {\bibfnamefont {M.}~\bibnamefont {Xi}}, \bibinfo {author} {\bibfnamefont {S.}~\bibnamefont {Tian}}, \bibinfo {author} {\bibfnamefont {Y.}~\bibnamefont {Li}}, \bibinfo {author} {\bibfnamefont {H.}~\bibnamefont {Lei}}, \bibinfo {author} {\bibfnamefont {Z.}~\bibnamefont {Luo}}, \bibinfo {author} {\bibfnamefont {J.}~\bibnamefont {Yang}}, \bibinfo {author} {\bibfnamefont {Y.}~\bibnamefont {Wang}}, \bibinfo {author} {\bibfnamefont {T.}~\bibnamefont {Xia}}, \bibinfo {author} {\bibfnamefont {W.}~\bibnamefont
  {Zhou}}, \bibinfo {author} {\bibfnamefont {Y.}~\bibnamefont {Huang}},\ and\ \bibinfo {author} {\bibfnamefont {Y.}~\bibnamefont {Ye}},\ }\bibinfo {title} {Ferromagnetic-antiferromagnetic coexisting ground state and exchange bias effects in {MnBi\textsubscript{4}Te\textsubscript{7}} and {MnBi\textsubscript{6}Te\textsubscript{10}}},\ \href {https://doi.org/10.1038/s41467-022-35184-7} {\bibfield  {journal} {\bibinfo  {journal} {Nature Communications}\ }\textbf {\bibinfo {volume} {13}},\ \bibinfo {pages} {7646} (\bibinfo {year} {2022}{\natexlab{a}})}\BibitemShut {NoStop}%
\bibitem [{\citenamefont {Ding}\ \emph {et~al.}(2020)\citenamefont {Ding}, \citenamefont {Wang}, \citenamefont {Xu},\ and\ \citenamefont {Li}}]{Ding2020}%
  \BibitemOpen
  \bibfield  {author} {\bibinfo {author} {\bibfnamefont {L.}~\bibnamefont {Ding}}, \bibinfo {author} {\bibfnamefont {X.}~\bibnamefont {Wang}}, \bibinfo {author} {\bibfnamefont {S.}~\bibnamefont {Xu}},\ and\ \bibinfo {author} {\bibfnamefont {Y.}~\bibnamefont {Li}},\ }\bibinfo {title} {Crystal and magnetic structures of {MnBi\textsubscript{4}Te\textsubscript{7}}: A layered magnetic topological material},\ \href {https://doi.org/10.1103/PhysRevB.101.020412} {\bibfield  {journal} {\bibinfo  {journal} {Phys. Rev. B}\ }\textbf {\bibinfo {volume} {101}},\ \bibinfo {pages} {020412} (\bibinfo {year} {2020})}\BibitemShut {NoStop}%
\bibitem [{\citenamefont {Yuan}\ \emph {et~al.}(2020)\citenamefont {Yuan}, \citenamefont {Wang}, \citenamefont {Li}, \citenamefont {Li}, \citenamefont {Ji}, \citenamefont {Hao}, \citenamefont {Wu}, \citenamefont {He}, \citenamefont {Wang}, \citenamefont {Xu} \emph {et~al.}}]{STMelectronic}%
  \BibitemOpen
  \bibfield  {author} {\bibinfo {author} {\bibfnamefont {Y.}~\bibnamefont {Yuan}}, \bibinfo {author} {\bibfnamefont {X.}~\bibnamefont {Wang}}, \bibinfo {author} {\bibfnamefont {H.}~\bibnamefont {Li}}, \bibinfo {author} {\bibfnamefont {J.}~\bibnamefont {Li}}, \bibinfo {author} {\bibfnamefont {Y.}~\bibnamefont {Ji}}, \bibinfo {author} {\bibfnamefont {Z.}~\bibnamefont {Hao}}, \bibinfo {author} {\bibfnamefont {Y.}~\bibnamefont {Wu}}, \bibinfo {author} {\bibfnamefont {K.}~\bibnamefont {He}}, \bibinfo {author} {\bibfnamefont {Y.}~\bibnamefont {Wang}}, \bibinfo {author} {\bibfnamefont {Y.}~\bibnamefont {Xu}}, \emph {et~al.},\ }\bibinfo {title} {Electronic states and magnetic response of {MnBi\textsubscript{2}Te\textsubscript{4}} by scanning tunneling microscopy and spectroscopy},\ \href {https://doi.org/10.1021/acs.nanolett.0c00031} {\bibfield  {journal} {\bibinfo  {journal} {Nano letters}\ }\textbf {\bibinfo {volume} {20}},\ \bibinfo {pages} {3271} (\bibinfo {year} {2020})}\BibitemShut {NoStop}%
\bibitem [{\citenamefont {Wu}\ \emph {et~al.}(2020)\citenamefont {Wu}, \citenamefont {Li}, \citenamefont {Ma}, \citenamefont {Zhang}, \citenamefont {Liu}, \citenamefont {Zhou}, \citenamefont {Shao}, \citenamefont {Wang}, \citenamefont {Hao}, \citenamefont {Feng} \emph {et~al.}}]{STMdistinct}%
  \BibitemOpen
  \bibfield  {author} {\bibinfo {author} {\bibfnamefont {X.}~\bibnamefont {Wu}}, \bibinfo {author} {\bibfnamefont {J.}~\bibnamefont {Li}}, \bibinfo {author} {\bibfnamefont {X.-M.}\ \bibnamefont {Ma}}, \bibinfo {author} {\bibfnamefont {Y.}~\bibnamefont {Zhang}}, \bibinfo {author} {\bibfnamefont {Y.}~\bibnamefont {Liu}}, \bibinfo {author} {\bibfnamefont {C.-S.}\ \bibnamefont {Zhou}}, \bibinfo {author} {\bibfnamefont {J.}~\bibnamefont {Shao}}, \bibinfo {author} {\bibfnamefont {Q.}~\bibnamefont {Wang}}, \bibinfo {author} {\bibfnamefont {Y.-J.}\ \bibnamefont {Hao}}, \bibinfo {author} {\bibfnamefont {Y.}~\bibnamefont {Feng}}, \emph {et~al.},\ }\bibinfo {title} {Distinct topological surface states on the two terminations of {MnBi\textsubscript{4}Te\textsubscript{7}}},\ \href {https://doi.org/10.1103/PhysRevX.10.031013} {\bibfield  {journal} {\bibinfo  {journal} {Physical Review X}\ }\textbf {\bibinfo {volume} {10}},\ \bibinfo {pages} {031013} (\bibinfo {year} {2020})}\BibitemShut {NoStop}%
\bibitem [{\citenamefont {Cui}\ \emph {et~al.}(2023)\citenamefont {Cui}, \citenamefont {Zhang},\ and\ \citenamefont {Wang}}]{Cui2023}%
  \BibitemOpen
  \bibfield  {author} {\bibinfo {author} {\bibfnamefont {J.}~\bibnamefont {Cui}}, \bibinfo {author} {\bibfnamefont {Y.}~\bibnamefont {Zhang}},\ and\ \bibinfo {author} {\bibfnamefont {Z.}~\bibnamefont {Wang}},\ }\bibinfo {title} {Layer-dependent magnetism and anomalous hall effect in {MnBi\textsubscript{4}Te\textsubscript{7}} thin flakes},\ \href {https://doi.org/10.1021/acs.nanolett.2c03773} {\bibfield  {journal} {\bibinfo  {journal} {Nano Lett.}\ }\textbf {\bibinfo {volume} {23}},\ \bibinfo {pages} {1234} (\bibinfo {year} {2023})}\BibitemShut {NoStop}%
\bibitem [{\citenamefont {Ge}\ \emph {et~al.}(2022)\citenamefont {Ge}, \citenamefont {Wu},\ and\ \citenamefont {Liu}}]{Ge2022}%
  \BibitemOpen
  \bibfield  {author} {\bibinfo {author} {\bibfnamefont {W.}~\bibnamefont {Ge}}, \bibinfo {author} {\bibfnamefont {H.}~\bibnamefont {Wu}},\ and\ \bibinfo {author} {\bibfnamefont {J.}~\bibnamefont {Liu}},\ }\bibinfo {title} {Surface spin-flip transitions in {MnBi\textsubscript{4}Te\textsubscript{7}} revealed by spectroscopy},\ \href {https://doi.org/10.1103/PhysRevLett.129.107204} {\bibfield  {journal} {\bibinfo  {journal} {Phys. Rev. Lett.}\ }\textbf {\bibinfo {volume} {129}},\ \bibinfo {pages} {107204} (\bibinfo {year} {2022})}\BibitemShut {NoStop}%
\bibitem [{\citenamefont {Kim}\ \emph {et~al.}(2024)\citenamefont {Kim}, \citenamefont {Lee},\ and\ \citenamefont {Kwon}}]{Kim2024}%
  \BibitemOpen
  \bibfield  {author} {\bibinfo {author} {\bibfnamefont {T.}~\bibnamefont {Kim}}, \bibinfo {author} {\bibfnamefont {S.}~\bibnamefont {Lee}},\ and\ \bibinfo {author} {\bibfnamefont {H.}~\bibnamefont {Kwon}},\ }\bibinfo {title} {Layer-controlled topological phase transitions in {MnBi\textsubscript{4}Te\textsubscript{7}}},\ \bibfield  {journal} {\bibinfo  {journal} {Nat. Commun.}\ }\textbf {\bibinfo {volume} {15}},\ \href {https://doi.org/10.1038/s41467-024-10234-5} {10.1038/s41467-024-10234-5} (\bibinfo {year} {2024})\BibitemShut {NoStop}%
\bibitem [{\citenamefont {Hu}\ \emph {et~al.}(2024)\citenamefont {Hu}, \citenamefont {Qian},\ and\ \citenamefont {Ni}}]{recentprogress}%
  \BibitemOpen
  \bibfield  {author} {\bibinfo {author} {\bibfnamefont {C.}~\bibnamefont {Hu}}, \bibinfo {author} {\bibfnamefont {T.}~\bibnamefont {Qian}},\ and\ \bibinfo {author} {\bibfnamefont {N.}~\bibnamefont {Ni}},\ }\bibinfo {title} {Recent progress in {MnBi\textsubscript{2n}Te\textsubscript{3n+1}} intrinsic magnetic topological insulators: Crystal growth, magnetism, and chemical disorder},\ \href@noop {} {\bibfield  {journal} {\bibinfo  {journal} {National Science Review}\ }\textbf {\bibinfo {volume} {11}},\ \bibinfo {pages} {nwad282} (\bibinfo {year} {2024})}\BibitemShut {NoStop}%
\bibitem [{\citenamefont {Guo}\ \emph {et~al.}(2024)\citenamefont {Guo}, \citenamefont {Wang}, \citenamefont {Zhang}, \citenamefont {Mi}, \citenamefont {Li}, \citenamefont {Dong}, \citenamefont {Zhu}, \citenamefont {Hu}, \citenamefont {Wang}, \citenamefont {Li} \emph {et~al.}}]{guo2024interlayer}%
  \BibitemOpen
  \bibfield  {author} {\bibinfo {author} {\bibfnamefont {J.}~\bibnamefont {Guo}}, \bibinfo {author} {\bibfnamefont {H.}~\bibnamefont {Wang}}, \bibinfo {author} {\bibfnamefont {H.}~\bibnamefont {Zhang}}, \bibinfo {author} {\bibfnamefont {S.}~\bibnamefont {Mi}}, \bibinfo {author} {\bibfnamefont {S.}~\bibnamefont {Li}}, \bibinfo {author} {\bibfnamefont {H.}~\bibnamefont {Dong}}, \bibinfo {author} {\bibfnamefont {S.}~\bibnamefont {Zhu}}, \bibinfo {author} {\bibfnamefont {J.}~\bibnamefont {Hu}}, \bibinfo {author} {\bibfnamefont {X.}~\bibnamefont {Wang}}, \bibinfo {author} {\bibfnamefont {Y.}~\bibnamefont {Li}}, \emph {et~al.},\ }\bibinfo {title} {Interlayer coupling modulated tunable magnetic states in superlattice {MnBi\textsubscript{2}Te\textsubscript{4}} ({Bi\textsubscript{2}Te\textsubscript{3}})\textsubscript{n} topological insulators},\ \href {https://doi.org/10.1103/PhysRevB.109.165410} {\bibfield  {journal} {\bibinfo  {journal} {Physical Review B}\ }\textbf {\bibinfo {volume} {109}},\ \bibinfo {pages}
  {165410} (\bibinfo {year} {2024})}\BibitemShut {NoStop}%
\bibitem [{\citenamefont {Shi}\ \emph {et~al.}(2019)\citenamefont {Shi}, \citenamefont {Lei}, \citenamefont {Zhu}, \citenamefont {Ma}, \citenamefont {Cui}, \citenamefont {Sun}, \citenamefont {Chen}, \citenamefont {Wang}, \citenamefont {Zhang}, \citenamefont {Li},\ and\ \citenamefont {Chen}}]{Shi2019}%
  \BibitemOpen
  \bibfield  {author} {\bibinfo {author} {\bibfnamefont {M.~Z.}\ \bibnamefont {Shi}}, \bibinfo {author} {\bibfnamefont {B.}~\bibnamefont {Lei}}, \bibinfo {author} {\bibfnamefont {C.~S.}\ \bibnamefont {Zhu}}, \bibinfo {author} {\bibfnamefont {D.~H.}\ \bibnamefont {Ma}}, \bibinfo {author} {\bibfnamefont {J.~H.}\ \bibnamefont {Cui}}, \bibinfo {author} {\bibfnamefont {Z.~L.}\ \bibnamefont {Sun}}, \bibinfo {author} {\bibfnamefont {H.}~\bibnamefont {Chen}}, \bibinfo {author} {\bibfnamefont {Y.~Y.}\ \bibnamefont {Wang}}, \bibinfo {author} {\bibfnamefont {T.}~\bibnamefont {Zhang}}, \bibinfo {author} {\bibfnamefont {Q.}~\bibnamefont {Li}},\ and\ \bibinfo {author} {\bibfnamefont {X.~H.}\ \bibnamefont {Chen}},\ }\bibinfo {title} {Magnetic and transport properties in the magnetic topological insulators {MnBi\textsubscript{2}Te\textsubscript{4}}({Bi\textsubscript{2}Te\textsubscript{3}})\textsubscript{n} (n = 1, 2)},\ \href {https://doi.org/10.1103/PhysRevB.100.155144} {\bibfield  {journal} {\bibinfo  {journal} {Physical
  Review B}\ }\textbf {\bibinfo {volume} {100}},\ \bibinfo {pages} {155144} (\bibinfo {year} {2019})}\BibitemShut {NoStop}%
\bibitem [{\citenamefont {Deng}\ \emph {et~al.}(2020)\citenamefont {Deng}, \citenamefont {Yu}, \citenamefont {Shi}, \citenamefont {Guo}, \citenamefont {Xu}, \citenamefont {Wang}, \citenamefont {Chen},\ and\ \citenamefont {Zhang}}]{Deng2020}%
  \BibitemOpen
  \bibfield  {author} {\bibinfo {author} {\bibfnamefont {Y.}~\bibnamefont {Deng}}, \bibinfo {author} {\bibfnamefont {Y.}~\bibnamefont {Yu}}, \bibinfo {author} {\bibfnamefont {M.-Z.}\ \bibnamefont {Shi}}, \bibinfo {author} {\bibfnamefont {Z.}~\bibnamefont {Guo}}, \bibinfo {author} {\bibfnamefont {Z.}~\bibnamefont {Xu}}, \bibinfo {author} {\bibfnamefont {J.}~\bibnamefont {Wang}}, \bibinfo {author} {\bibfnamefont {X.}~\bibnamefont {Chen}},\ and\ \bibinfo {author} {\bibfnamefont {Y.}~\bibnamefont {Zhang}},\ }\bibinfo {title} {Quantum anomalous hall effect in intrinsic magnetic topological insulator {MnBi\textsubscript{2}Te\textsubscript{4}}},\ \href {https://doi.org/10.1126/science.aax8156} {\bibfield  {journal} {\bibinfo  {journal} {Science}\ }\textbf {\bibinfo {volume} {367}},\ \bibinfo {pages} {895} (\bibinfo {year} {2020})}\BibitemShut {NoStop}%
\bibitem [{\citenamefont {Zhang}\ \emph {et~al.}(2019)\citenamefont {Zhang}, \citenamefont {Shi}, \citenamefont {Zhu}, \citenamefont {Xing}, \citenamefont {Zhang},\ and\ \citenamefont {Wang}}]{axion}%
  \BibitemOpen
  \bibfield  {author} {\bibinfo {author} {\bibfnamefont {D.}~\bibnamefont {Zhang}}, \bibinfo {author} {\bibfnamefont {M.}~\bibnamefont {Shi}}, \bibinfo {author} {\bibfnamefont {T.}~\bibnamefont {Zhu}}, \bibinfo {author} {\bibfnamefont {D.}~\bibnamefont {Xing}}, \bibinfo {author} {\bibfnamefont {H.}~\bibnamefont {Zhang}},\ and\ \bibinfo {author} {\bibfnamefont {J.}~\bibnamefont {Wang}},\ }\bibinfo {title} {Topological axion states in the magnetic insulator {MnBi\textsubscript{2}Te\textsubscript{4}} with the quantized magnetoelectric effect},\ \href {https://doi.org/10.1103/PhysRevLett.122.206401} {\bibfield  {journal} {\bibinfo  {journal} {Physical review letters}\ }\textbf {\bibinfo {volume} {122}},\ \bibinfo {pages} {206401} (\bibinfo {year} {2019})}\BibitemShut {NoStop}%
\bibitem [{\citenamefont {Gibertini}\ \emph {et~al.}(2019)\citenamefont {Gibertini}, \citenamefont {Koperski}, \citenamefont {Morpurgo},\ and\ \citenamefont {Novoselov}}]{Gibertini2019Magnetic2D}%
  \BibitemOpen
  \bibfield  {author} {\bibinfo {author} {\bibfnamefont {M.}~\bibnamefont {Gibertini}}, \bibinfo {author} {\bibfnamefont {M.}~\bibnamefont {Koperski}}, \bibinfo {author} {\bibfnamefont {A.~F.}\ \bibnamefont {Morpurgo}},\ and\ \bibinfo {author} {\bibfnamefont {K.~S.}\ \bibnamefont {Novoselov}},\ }\bibinfo {title} {Magnetic 2d materials and heterostructures},\ \href {https://doi.org/10.1038/s41565-019-0438-6} {\bibfield  {journal} {\bibinfo  {journal} {Nature Nanotechnology}\ }\textbf {\bibinfo {volume} {14}},\ \bibinfo {pages} {408} (\bibinfo {year} {2019})}\BibitemShut {NoStop}%
\bibitem [{\citenamefont {Gao}\ \emph {et~al.}(2021)\citenamefont {Gao}, \citenamefont {Liu}, \citenamefont {Hu}, \citenamefont {Qiu}, \citenamefont {Tzschaschel}, \citenamefont {Ghosh}, \citenamefont {Liu}, \citenamefont {Liu}, \citenamefont {Liu}, \citenamefont {Liu}, \citenamefont {Tang}, \citenamefont {Wang}, \citenamefont {Wang}, \citenamefont {Ma}, \citenamefont {Liu}, \citenamefont {Jarillo-Herrero}, \citenamefont {Gedik}, \citenamefont {Liu}, \citenamefont {Chang},\ and\ \citenamefont {Xu}}]{layerhalleffect}%
  \BibitemOpen
  \bibfield  {author} {\bibinfo {author} {\bibfnamefont {A.}~\bibnamefont {Gao}}, \bibinfo {author} {\bibfnamefont {Y.~F.}\ \bibnamefont {Liu}}, \bibinfo {author} {\bibfnamefont {C.}~\bibnamefont {Hu}}, \bibinfo {author} {\bibfnamefont {J.~X.}\ \bibnamefont {Qiu}}, \bibinfo {author} {\bibfnamefont {C.}~\bibnamefont {Tzschaschel}}, \bibinfo {author} {\bibfnamefont {B.}~\bibnamefont {Ghosh}}, \bibinfo {author} {\bibfnamefont {P.}~\bibnamefont {Liu}}, \bibinfo {author} {\bibfnamefont {J.}~\bibnamefont {Liu}}, \bibinfo {author} {\bibfnamefont {Y.}~\bibnamefont {Liu}}, \bibinfo {author} {\bibfnamefont {Y.}~\bibnamefont {Liu}}, \bibinfo {author} {\bibfnamefont {R.}~\bibnamefont {Tang}}, \bibinfo {author} {\bibfnamefont {S.}~\bibnamefont {Wang}}, \bibinfo {author} {\bibfnamefont {X.}~\bibnamefont {Wang}}, \bibinfo {author} {\bibfnamefont {Q.}~\bibnamefont {Ma}}, \bibinfo {author} {\bibfnamefont {C.}~\bibnamefont {Liu}}, \bibinfo {author} {\bibfnamefont {P.}~\bibnamefont {Jarillo-Herrero}}, \bibinfo {author}
  {\bibfnamefont {N.}~\bibnamefont {Gedik}}, \bibinfo {author} {\bibfnamefont {Q.}~\bibnamefont {Liu}}, \bibinfo {author} {\bibfnamefont {T.-R.}\ \bibnamefont {Chang}},\ and\ \bibinfo {author} {\bibfnamefont {S.~Y.}\ \bibnamefont {Xu}},\ }\bibinfo {title} {Layer hall effect in a 2d topological axion antiferromagnet},\ \href {https://doi.org/10.1038/s41586-021-03679-w} {\bibfield  {journal} {\bibinfo  {journal} {Nature}\ }\textbf {\bibinfo {volume} {595}},\ \bibinfo {pages} {521} (\bibinfo {year} {2021})}\BibitemShut {NoStop}%
\bibitem [{\citenamefont {Zhao}\ and\ \citenamefont {Liu}(2021)}]{Zhao2021}%
  \BibitemOpen
  \bibfield  {author} {\bibinfo {author} {\bibfnamefont {Y.}~\bibnamefont {Zhao}}\ and\ \bibinfo {author} {\bibfnamefont {Q.}~\bibnamefont {Liu}},\ }\bibinfo {title} {Routes to realize the axion-insulator phase in {MnBi\textsubscript{2}Te\textsubscript{4}}({Bi\textsubscript{2}Te\textsubscript{3}})\textsubscript{n} family},\ \href {https://doi.org/10.1063/5.0059447} {\bibfield  {journal} {\bibinfo  {journal} {Applied Physics Letters}\ }\textbf {\bibinfo {volume} {119}},\ \bibinfo {pages} {060502} (\bibinfo {year} {2021})}\BibitemShut {NoStop}%
\bibitem [{\citenamefont {Li}\ \emph {et~al.}(2024)\citenamefont {Li}, \citenamefont {Gong}, \citenamefont {Cheng}, \citenamefont {Jiang},\ and\ \citenamefont {Xie}}]{layertronics}%
  \BibitemOpen
  \bibfield  {author} {\bibinfo {author} {\bibfnamefont {S.}~\bibnamefont {Li}}, \bibinfo {author} {\bibfnamefont {M.}~\bibnamefont {Gong}}, \bibinfo {author} {\bibfnamefont {S.}~\bibnamefont {Cheng}}, \bibinfo {author} {\bibfnamefont {H.}~\bibnamefont {Jiang}},\ and\ \bibinfo {author} {\bibfnamefont {X.~C.}\ \bibnamefont {Xie}},\ }\bibinfo {title} {Dissipationless layertronics in axion insulator {MnBi\textsubscript{2}Te\textsubscript{4}}},\ \href {https://doi.org/10.1093/nsr/nwad262} {\bibfield  {journal} {\bibinfo  {journal} {National Science Review}\ }\textbf {\bibinfo {volume} {11}},\ \bibinfo {pages} {nwad262} (\bibinfo {year} {2024})}\BibitemShut {NoStop}%
\bibitem [{\citenamefont {Zhang}\ \emph {et~al.}(2024)\citenamefont {Zhang}, \citenamefont {Wu}, \citenamefont {Liu}, \citenamefont {Huang}, \citenamefont {Gao}, \citenamefont {Yang}, \citenamefont {Zhao}, \citenamefont {Li}, \citenamefont {Ma},\ and\ \citenamefont {Zhang}}]{zhang2024metamagnetic}%
  \BibitemOpen
  \bibfield  {author} {\bibinfo {author} {\bibfnamefont {H.}~\bibnamefont {Zhang}}, \bibinfo {author} {\bibfnamefont {H.}~\bibnamefont {Wu}}, \bibinfo {author} {\bibfnamefont {D.}~\bibnamefont {Liu}}, \bibinfo {author} {\bibfnamefont {J.}~\bibnamefont {Huang}}, \bibinfo {author} {\bibfnamefont {F.}~\bibnamefont {Gao}}, \bibinfo {author} {\bibfnamefont {T.}~\bibnamefont {Yang}}, \bibinfo {author} {\bibfnamefont {X.}~\bibnamefont {Zhao}}, \bibinfo {author} {\bibfnamefont {B.}~\bibnamefont {Li}}, \bibinfo {author} {\bibfnamefont {S.}~\bibnamefont {Ma}},\ and\ \bibinfo {author} {\bibfnamefont {Z.}~\bibnamefont {Zhang}},\ }\bibinfo {title} {Metamagnetic tricritical behavior of the magnetic topological insulator {MnBi\textsubscript{4}Te\textsubscript{7}}},\ \href {https://doi.org/10.1103/PhysRevB.109.214428} {\bibfield  {journal} {\bibinfo  {journal} {Physical Review B}\ }\textbf {\bibinfo {volume} {109}},\ \bibinfo {pages} {214428} (\bibinfo {year} {2024})}\BibitemShut {NoStop}%
\bibitem [{\citenamefont {Das}\ and\ \citenamefont {Mukhopadhyay}(2025)}]{das2025metamagnetic}%
  \BibitemOpen
  \bibfield  {author} {\bibinfo {author} {\bibfnamefont {T.}~\bibnamefont {Das}}\ and\ \bibinfo {author} {\bibfnamefont {S.}~\bibnamefont {Mukhopadhyay}},\ }\bibinfo {title} {Metamagnetic quantum criticality in the antiferromagnetic topological insulator {MnBi\textsubscript{2}Te\textsubscript{4}}},\ \href {https://doi.org/10.1103/PhysRevB.111.174420} {\bibfield  {journal} {\bibinfo  {journal} {Physical Review B}\ }\textbf {\bibinfo {volume} {111}},\ \bibinfo {pages} {174420} (\bibinfo {year} {2025})}\BibitemShut {NoStop}%
\bibitem [{\citenamefont {Firdosh}\ \emph {et~al.}(2025)\citenamefont {Firdosh}, \citenamefont {Sinha}, \citenamefont {Sinha}, \citenamefont {Singh}, \citenamefont {Patnaik},\ and\ \citenamefont {Manna}}]{firdosh2025exchange}%
  \BibitemOpen
  \bibfield  {author} {\bibinfo {author} {\bibfnamefont {N.}~\bibnamefont {Firdosh}}, \bibinfo {author} {\bibfnamefont {S.}~\bibnamefont {Sinha}}, \bibinfo {author} {\bibfnamefont {I.}~\bibnamefont {Sinha}}, \bibinfo {author} {\bibfnamefont {M.}~\bibnamefont {Singh}}, \bibinfo {author} {\bibfnamefont {S.}~\bibnamefont {Patnaik}},\ and\ \bibinfo {author} {\bibfnamefont {S.}~\bibnamefont {Manna}},\ }\bibinfo {title} {Exchange bias and anomalous hall effect driven by intertwined magnetic phases in {MnBi\textsubscript{4}Te\textsubscript{7}}},\ \href {https://doi.org/10.1103/pjh8-j9p1} {\bibfield  {journal} {\bibinfo  {journal} {Physical Review B}\ }\textbf {\bibinfo {volume} {112}},\ \bibinfo {pages} {144410} (\bibinfo {year} {2025})}\BibitemShut {NoStop}%
\bibitem [{\citenamefont {Yang}\ \emph {et~al.}(2024)\citenamefont {Yang}, \citenamefont {Pan}, \citenamefont {He},\ and\ \citenamefont {Chu}}]{criticalMnSb}%
  \BibitemOpen
  \bibfield  {author} {\bibinfo {author} {\bibfnamefont {X.}~\bibnamefont {Yang}}, \bibinfo {author} {\bibfnamefont {J.}~\bibnamefont {Pan}}, \bibinfo {author} {\bibfnamefont {X.}~\bibnamefont {He}},\ and\ \bibinfo {author} {\bibfnamefont {D.}~\bibnamefont {Chu}},\ }\bibinfo {title} {Critical behavior, magnetic phase diagram, and magnetic entropy change of {MnSb\textsubscript{4}Te\textsubscript{7}}},\ \href {https://doi.org/10.1103/PhysRevB.109.094408} {\bibfield  {journal} {\bibinfo  {journal} {Physical Review B}\ }\textbf {\bibinfo {volume} {109}},\ \bibinfo {pages} {094408} (\bibinfo {year} {2024})}\BibitemShut {NoStop}%
\bibitem [{\citenamefont {Ding}\ \emph {et~al.}(2021)\citenamefont {Ding}, \citenamefont {Hu}, \citenamefont {Feng}, \citenamefont {Jiang}, \citenamefont {Kibalin}, \citenamefont {Gukasov},\ and\ \citenamefont {Cao}}]{NeutronMBT}%
  \BibitemOpen
  \bibfield  {author} {\bibinfo {author} {\bibfnamefont {L.}~\bibnamefont {Ding}}, \bibinfo {author} {\bibfnamefont {C.}~\bibnamefont {Hu}}, \bibinfo {author} {\bibfnamefont {E.}~\bibnamefont {Feng}}, \bibinfo {author} {\bibfnamefont {C.}~\bibnamefont {Jiang}}, \bibinfo {author} {\bibfnamefont {I.~A.}\ \bibnamefont {Kibalin}}, \bibinfo {author} {\bibfnamefont {A.}~\bibnamefont {Gukasov}},\ and\ \bibinfo {author} {\bibfnamefont {H.}~\bibnamefont {Cao}},\ }\bibinfo {title} {Neutron diffraction study of magnetism in van der waals layered {MnBi\textsubscript{2n}Te\textsubscript{3n+1}}},\ \href {https://doi.org/10.1088/1361-6463/abe0dd} {\bibfield  {journal} {\bibinfo  {journal} {Journal of Physics D: Applied Physics}\ }\textbf {\bibinfo {volume} {54}},\ \bibinfo {pages} {174003} (\bibinfo {year} {2021})}\BibitemShut {NoStop}%
\bibitem [{\citenamefont {Sun}\ \emph {et~al.}(2023)\citenamefont {Sun}, \citenamefont {Zhang},\ and\ \citenamefont {Li}}]{Sun2023}%
  \BibitemOpen
  \bibfield  {author} {\bibinfo {author} {\bibfnamefont {Y.}~\bibnamefont {Sun}}, \bibinfo {author} {\bibfnamefont {L.}~\bibnamefont {Zhang}},\ and\ \bibinfo {author} {\bibfnamefont {H.}~\bibnamefont {Li}},\ }\bibinfo {title} {Axion electrodynamics in antiferromagnetic topological insulators},\ \href {https://doi.org/10.1103/PhysRevMaterials.7.074204} {\bibfield  {journal} {\bibinfo  {journal} {Phys. Rev. Materials}\ }\textbf {\bibinfo {volume} {7}},\ \bibinfo {pages} {074204} (\bibinfo {year} {2023})}\BibitemShut {NoStop}%
\bibitem [{\citenamefont {He}\ \emph {et~al.}(2022)\citenamefont {He}, \citenamefont {Li},\ and\ \citenamefont {Xu}}]{He2022}%
  \BibitemOpen
  \bibfield  {author} {\bibinfo {author} {\bibfnamefont {Q.}~\bibnamefont {He}}, \bibinfo {author} {\bibfnamefont {M.}~\bibnamefont {Li}},\ and\ \bibinfo {author} {\bibfnamefont {X.}~\bibnamefont {Xu}},\ }\bibinfo {title} {Van der waals magnets for next-generation quantum technologies},\ \href {https://doi.org/10.1039/D2TC01847A} {\bibfield  {journal} {\bibinfo  {journal} {J. Mater. Chem. C}\ }\textbf {\bibinfo {volume} {10}},\ \bibinfo {pages} {13214} (\bibinfo {year} {2022})}\BibitemShut {NoStop}%
\bibitem [{\citenamefont {Klimovskikh}\ \emph {et~al.}(2020)\citenamefont {Klimovskikh}, \citenamefont {Otrokov}, \citenamefont {Estyunin}, \citenamefont {Eremeev}, \citenamefont {Filnov}, \citenamefont {Koroleva}, \citenamefont {Isaeva}, \citenamefont {Kißner}, \citenamefont {Ünzelmann}, \citenamefont {Min}, \citenamefont {Moser}, \citenamefont {Peixoto}, \citenamefont {Reinert}, \citenamefont {Ernst}, \citenamefont {Echenique},\ and\ \citenamefont {Chulkov}}]{Klimovskikh2020TunableMagnetism}%
  \BibitemOpen
  \bibfield  {author} {\bibinfo {author} {\bibfnamefont {I.~I.}\ \bibnamefont {Klimovskikh}}, \bibinfo {author} {\bibfnamefont {M.~M.}\ \bibnamefont {Otrokov}}, \bibinfo {author} {\bibfnamefont {D.}~\bibnamefont {Estyunin}}, \bibinfo {author} {\bibfnamefont {S.~V.}\ \bibnamefont {Eremeev}}, \bibinfo {author} {\bibfnamefont {S.~O.}\ \bibnamefont {Filnov}}, \bibinfo {author} {\bibfnamefont {A.}~\bibnamefont {Koroleva}}, \bibinfo {author} {\bibfnamefont {A.}~\bibnamefont {Isaeva}}, \bibinfo {author} {\bibfnamefont {K.}~\bibnamefont {Kißner}}, \bibinfo {author} {\bibfnamefont {M.}~\bibnamefont {Ünzelmann}}, \bibinfo {author} {\bibfnamefont {C.~H.}\ \bibnamefont {Min}}, \bibinfo {author} {\bibfnamefont {S.}~\bibnamefont {Moser}}, \bibinfo {author} {\bibfnamefont {T.~R.~F.}\ \bibnamefont {Peixoto}}, \bibinfo {author} {\bibfnamefont {F.}~\bibnamefont {Reinert}}, \bibinfo {author} {\bibfnamefont {A.}~\bibnamefont {Ernst}}, \bibinfo {author} {\bibfnamefont {P.~M.}\ \bibnamefont {Echenique}},\ and\ \bibinfo {author}
  {\bibfnamefont {E.~V.}\ \bibnamefont {Chulkov}},\ }\bibinfo {title} {Tunable 3d/2d magnetism in the {MnBi\textsubscript{2}Te\textsubscript{4}} ({Bi\textsubscript{2}Te\textsubscript{3}})\textsubscript{m} topological insulators family},\ \href {https://doi.org/10.1038/s41535-020-00255-9} {\bibfield  {journal} {\bibinfo  {journal} {npj Quantum Materials}\ }\textbf {\bibinfo {volume} {5}},\ \bibinfo {pages} {54} (\bibinfo {year} {2020})}\BibitemShut {NoStop}%
\bibitem [{\citenamefont {Xu}\ \emph {et~al.}(2022{\natexlab{b}})\citenamefont {Xu}, \citenamefont {Gu}, \citenamefont {Fei}, \citenamefont {Gu}, \citenamefont {Liu}, \citenamefont {Yu}, \citenamefont {Xue}, \citenamefont {Ning}, \citenamefont {Chen}, \citenamefont {Xie} \emph {et~al.}}]{xuedgeobservation}%
  \BibitemOpen
  \bibfield  {author} {\bibinfo {author} {\bibfnamefont {H.-K.}\ \bibnamefont {Xu}}, \bibinfo {author} {\bibfnamefont {M.}~\bibnamefont {Gu}}, \bibinfo {author} {\bibfnamefont {F.}~\bibnamefont {Fei}}, \bibinfo {author} {\bibfnamefont {Y.-S.}\ \bibnamefont {Gu}}, \bibinfo {author} {\bibfnamefont {D.}~\bibnamefont {Liu}}, \bibinfo {author} {\bibfnamefont {Q.-Y.}\ \bibnamefont {Yu}}, \bibinfo {author} {\bibfnamefont {S.-S.}\ \bibnamefont {Xue}}, \bibinfo {author} {\bibfnamefont {X.-H.}\ \bibnamefont {Ning}}, \bibinfo {author} {\bibfnamefont {B.}~\bibnamefont {Chen}}, \bibinfo {author} {\bibfnamefont {H.}~\bibnamefont {Xie}}, \emph {et~al.},\ }\bibinfo {title} {Observation of magnetism-induced topological edge state in antiferromagnetic topological insulator {MnBi\textsubscript{4}Te\textsubscript{7}}},\ \href {https://doi.org/10.1021/acsnano.2c03622} {\bibfield  {journal} {\bibinfo  {journal} {ACS nano}\ }\textbf {\bibinfo {volume} {16}},\ \bibinfo {pages} {9810} (\bibinfo {year}
  {2022}{\natexlab{b}})}\BibitemShut {NoStop}%
\bibitem [{\citenamefont {Vidal}\ \emph {et~al.}(2019)\citenamefont {Vidal}, \citenamefont {Zeugner}, \citenamefont {Facio}, \citenamefont {Ray}, \citenamefont {Haghighi}, \citenamefont {Wolter}, \citenamefont {Corredor~Bohorquez}, \citenamefont {Caglieris}, \citenamefont {Moser}, \citenamefont {Figgemeier} \emph {et~al.}}]{vidal2019topological}%
  \BibitemOpen
  \bibfield  {author} {\bibinfo {author} {\bibfnamefont {R.~C.}\ \bibnamefont {Vidal}}, \bibinfo {author} {\bibfnamefont {A.}~\bibnamefont {Zeugner}}, \bibinfo {author} {\bibfnamefont {J.~I.}\ \bibnamefont {Facio}}, \bibinfo {author} {\bibfnamefont {R.}~\bibnamefont {Ray}}, \bibinfo {author} {\bibfnamefont {M.~H.}\ \bibnamefont {Haghighi}}, \bibinfo {author} {\bibfnamefont {A.~U.}\ \bibnamefont {Wolter}}, \bibinfo {author} {\bibfnamefont {L.~T.}\ \bibnamefont {Corredor~Bohorquez}}, \bibinfo {author} {\bibfnamefont {F.}~\bibnamefont {Caglieris}}, \bibinfo {author} {\bibfnamefont {S.}~\bibnamefont {Moser}}, \bibinfo {author} {\bibfnamefont {T.}~\bibnamefont {Figgemeier}}, \emph {et~al.},\ }\bibinfo {title} {Topological electronic structure and intrinsic magnetization in {MnBi\textsubscript{4}Te\textsubscript{7}}: a {Bi\textsubscript{2}Te\textsubscript{3}} derivative with a periodic mn sublattice},\ \href {https://doi.org/10.1103/PhysRevX.9.041065} {\bibfield  {journal} {\bibinfo  {journal} {Physical Review X}\
  }\textbf {\bibinfo {volume} {9}},\ \bibinfo {pages} {041065} (\bibinfo {year} {2019})}\BibitemShut {NoStop}%
\bibitem [{\citenamefont {Cui}\ \emph {et~al.}(2012)\citenamefont {Cui}, \citenamefont {Liu}, \citenamefont {Gong}, \citenamefont {Liu}, \citenamefont {Guo}, \citenamefont {Yang},\ and\ \citenamefont {Zhang}}]{nonmagneticSpacer}%
  \BibitemOpen
  \bibfield  {author} {\bibinfo {author} {\bibfnamefont {W.~B.}\ \bibnamefont {Cui}}, \bibinfo {author} {\bibfnamefont {W.}~\bibnamefont {Liu}}, \bibinfo {author} {\bibfnamefont {W.~J.}\ \bibnamefont {Gong}}, \bibinfo {author} {\bibfnamefont {X.~H.}\ \bibnamefont {Liu}}, \bibinfo {author} {\bibfnamefont {S.}~\bibnamefont {Guo}}, \bibinfo {author} {\bibfnamefont {F.}~\bibnamefont {Yang}},\ and\ \bibinfo {author} {\bibfnamefont {Z.~D.}\ \bibnamefont {Zhang}},\ }\bibinfo {title} {Exchange coupling in hard/soft-magnetic multilayer films with non-magnetic spacer layers},\ \href {https://doi.org/10.1063/1.3671774} {\bibfield  {journal} {\bibinfo  {journal} {Journal of Applied Physics}\ }\textbf {\bibinfo {volume} {111}},\ \bibinfo {pages} {07B533} (\bibinfo {year} {2012})}\BibitemShut {NoStop}%
\bibitem [{\citenamefont {Li}\ \emph {et~al.}(2025)\citenamefont {Li}, \citenamefont {Pajerowski}, \citenamefont {Yan},\ and\ \citenamefont {McQueeney}}]{SpacerMBT}%
  \BibitemOpen
  \bibfield  {author} {\bibinfo {author} {\bibfnamefont {B.}~\bibnamefont {Li}}, \bibinfo {author} {\bibfnamefont {D.~M.}\ \bibnamefont {Pajerowski}}, \bibinfo {author} {\bibfnamefont {J.~Q.}\ \bibnamefont {Yan}},\ and\ \bibinfo {author} {\bibfnamefont {R.~J.}\ \bibnamefont {McQueeney}},\ }\bibinfo {title} {Role of nonmagnetic spacers in the magnetic interactions of antiferromagnetic topological insulators {MnBi\textsubscript{4}Te\textsubscript{7}} and {MnBi\textsubscript{2}Te\textsubscript{4}}},\ \href {https://doi.org/10.1103/PhysRevB.111.064418} {\bibfield  {journal} {\bibinfo  {journal} {Physical Review B}\ }\textbf {\bibinfo {volume} {111}},\ \bibinfo {pages} {064418} (\bibinfo {year} {2025})}\BibitemShut {NoStop}%
\bibitem [{\citenamefont {Yang}\ \emph {et~al.}(2023)\citenamefont {Yang}, \citenamefont {Wu},\ and\ \citenamefont {Liu}}]{Yang2023}%
  \BibitemOpen
  \bibfield  {author} {\bibinfo {author} {\bibfnamefont {L.}~\bibnamefont {Yang}}, \bibinfo {author} {\bibfnamefont {J.}~\bibnamefont {Wu}},\ and\ \bibinfo {author} {\bibfnamefont {Q.}~\bibnamefont {Liu}},\ }\bibinfo {title} {High-temperature chern insulators in layered {MnBi\textsubscript{4}Te\textsubscript{7}}},\ \href {https://doi.org/10.1002/aelm.202201280} {\bibfield  {journal} {\bibinfo  {journal} {Adv. Electron. Mater.}\ }\textbf {\bibinfo {volume} {9}},\ \bibinfo {pages} {2201280} (\bibinfo {year} {2023})}\BibitemShut {NoStop}%
\bibitem [{\citenamefont {Zhao}\ \emph {et~al.}(2023)\citenamefont {Zhao}, \citenamefont {Liu},\ and\ \citenamefont {Wang}}]{Zhao2023}%
  \BibitemOpen
  \bibfield  {author} {\bibinfo {author} {\bibfnamefont {B.}~\bibnamefont {Zhao}}, \bibinfo {author} {\bibfnamefont {Y.}~\bibnamefont {Liu}},\ and\ \bibinfo {author} {\bibfnamefont {P.}~\bibnamefont {Wang}},\ }\bibinfo {title} {Manipulating spin textures in layered magnetic topological insulators},\ \href {https://doi.org/10.1103/PhysRevB.108.094423} {\bibfield  {journal} {\bibinfo  {journal} {Phys. Rev. B}\ }\textbf {\bibinfo {volume} {108}},\ \bibinfo {pages} {094423} (\bibinfo {year} {2023})}\BibitemShut {NoStop}%
\bibitem [{\citenamefont {Fisher}(1967)}]{fisher1967theory}%
  \BibitemOpen
  \bibfield  {author} {\bibinfo {author} {\bibfnamefont {M.~E.}\ \bibnamefont {Fisher}},\ }\bibinfo {title} {The theory of equilibrium critical phenomena},\ \href {https://doi.org/10.1088/0034-4885/30/2/306} {\bibfield  {journal} {\bibinfo  {journal} {Reports on progress in physics}\ }\textbf {\bibinfo {volume} {30}},\ \bibinfo {pages} {615} (\bibinfo {year} {1967})}\BibitemShut {NoStop}%
\bibitem [{\citenamefont {Fisher}\ \emph {et~al.}(1972)\citenamefont {Fisher}, \citenamefont {Ma},\ and\ \citenamefont {Nickel}}]{fisher1972critical}%
  \BibitemOpen
  \bibfield  {author} {\bibinfo {author} {\bibfnamefont {M.~E.}\ \bibnamefont {Fisher}}, \bibinfo {author} {\bibfnamefont {S.-k.}\ \bibnamefont {Ma}},\ and\ \bibinfo {author} {\bibfnamefont {B.}~\bibnamefont {Nickel}},\ }\bibinfo {title} {Critical exponents for long-range interactions},\ \href {https://doi.org/10.1103/PhysRevLett.29.917} {\bibfield  {journal} {\bibinfo  {journal} {Physical Review Letters}\ }\textbf {\bibinfo {volume} {29}},\ \bibinfo {pages} {917} (\bibinfo {year} {1972})}\BibitemShut {NoStop}%
\bibitem [{\citenamefont {Franco}\ and\ \citenamefont {Conde}(2010)}]{entropyrcpthetan}%
  \BibitemOpen
  \bibfield  {author} {\bibinfo {author} {\bibfnamefont {V.}~\bibnamefont {Franco}}\ and\ \bibinfo {author} {\bibfnamefont {A.}~\bibnamefont {Conde}},\ }\bibinfo {title} {Scaling laws for the magnetocaloric effect in second order phase transitions: From physics to applications for the characterization of materials},\ \href {https://doi.org/10.1016/j.ijrefrig.2009.12.019} {\bibfield  {journal} {\bibinfo  {journal} {international journal of refrigeration}\ }\textbf {\bibinfo {volume} {33}},\ \bibinfo {pages} {465} (\bibinfo {year} {2010})}\BibitemShut {NoStop}%
\bibitem [{\citenamefont {Chauhan}\ \emph {et~al.}(2022)\citenamefont {Chauhan}, \citenamefont {Kumar}, \citenamefont {Tiwari}, \citenamefont {Tiwari},\ and\ \citenamefont {Ghosh}}]{chauhan2022different}%
  \BibitemOpen
  \bibfield  {author} {\bibinfo {author} {\bibfnamefont {H.~C.}\ \bibnamefont {Chauhan}}, \bibinfo {author} {\bibfnamefont {B.}~\bibnamefont {Kumar}}, \bibinfo {author} {\bibfnamefont {A.}~\bibnamefont {Tiwari}}, \bibinfo {author} {\bibfnamefont {J.~K.}\ \bibnamefont {Tiwari}},\ and\ \bibinfo {author} {\bibfnamefont {S.}~\bibnamefont {Ghosh}},\ }\bibinfo {title} {Different critical exponents on two sides of a transition: Observation of crossover from ising to heisenberg exchange in skyrmion host {Cu\textsubscript{2}OSeO\textsubscript{3}}},\ \href {https://doi.org/10.1103/PhysRevLett.128.015703} {\bibfield  {journal} {\bibinfo  {journal} {Physical Review Letters}\ }\textbf {\bibinfo {volume} {128}},\ \bibinfo {pages} {015703} (\bibinfo {year} {2022})}\BibitemShut {NoStop}%
\bibitem [{\citenamefont {Zhang}\ \emph {et~al.}(2016)\citenamefont {Zhang}, \citenamefont {Han}, \citenamefont {Ge}, \citenamefont {Du}, \citenamefont {Jin}, \citenamefont {Wei}, \citenamefont {Fan}, \citenamefont {Zhang}, \citenamefont {Pi},\ and\ \citenamefont {Zhang}}]{zhang2016critical}%
  \BibitemOpen
  \bibfield  {author} {\bibinfo {author} {\bibfnamefont {L.}~\bibnamefont {Zhang}}, \bibinfo {author} {\bibfnamefont {H.}~\bibnamefont {Han}}, \bibinfo {author} {\bibfnamefont {M.}~\bibnamefont {Ge}}, \bibinfo {author} {\bibfnamefont {H.}~\bibnamefont {Du}}, \bibinfo {author} {\bibfnamefont {C.}~\bibnamefont {Jin}}, \bibinfo {author} {\bibfnamefont {W.}~\bibnamefont {Wei}}, \bibinfo {author} {\bibfnamefont {J.}~\bibnamefont {Fan}}, \bibinfo {author} {\bibfnamefont {C.}~\bibnamefont {Zhang}}, \bibinfo {author} {\bibfnamefont {L.}~\bibnamefont {Pi}},\ and\ \bibinfo {author} {\bibfnamefont {Y.}~\bibnamefont {Zhang}},\ }\bibinfo {title} {Critical phenomenon of the near room temperature skyrmion material fege},\ \href@noop {} {\bibfield  {journal} {\bibinfo  {journal} {Scientific reports}\ }\textbf {\bibinfo {volume} {6}},\ \bibinfo {pages} {22397} (\bibinfo {year} {2016})}\BibitemShut {NoStop}%
\bibitem [{\citenamefont {Meng}\ \emph {et~al.}(2023)\citenamefont {Meng}, \citenamefont {Liu}, \citenamefont {Rahman}, \citenamefont {Zhang}, \citenamefont {Fan}, \citenamefont {Ma}, \citenamefont {Ge}, \citenamefont {Yao}, \citenamefont {Pi}, \citenamefont {Zhang} \emph {et~al.}}]{meng2023crossover}%
  \BibitemOpen
  \bibfield  {author} {\bibinfo {author} {\bibfnamefont {F.}~\bibnamefont {Meng}}, \bibinfo {author} {\bibfnamefont {W.}~\bibnamefont {Liu}}, \bibinfo {author} {\bibfnamefont {A.}~\bibnamefont {Rahman}}, \bibinfo {author} {\bibfnamefont {J.}~\bibnamefont {Zhang}}, \bibinfo {author} {\bibfnamefont {J.}~\bibnamefont {Fan}}, \bibinfo {author} {\bibfnamefont {C.}~\bibnamefont {Ma}}, \bibinfo {author} {\bibfnamefont {M.}~\bibnamefont {Ge}}, \bibinfo {author} {\bibfnamefont {T.}~\bibnamefont {Yao}}, \bibinfo {author} {\bibfnamefont {L.}~\bibnamefont {Pi}}, \bibinfo {author} {\bibfnamefont {L.}~\bibnamefont {Zhang}}, \emph {et~al.},\ }\bibinfo {title} {Crossover of critical behavior and nontrivial magnetism in the chiral soliton lattice host {Cr\textsubscript{1/3}TaS\textsubscript{2}}},\ \href {https://doi.org/10.1103/PhysRevB.107.144425} {\bibfield  {journal} {\bibinfo  {journal} {Physical Review B}\ }\textbf {\bibinfo {volume} {107}},\ \bibinfo {pages} {144425} (\bibinfo {year} {2023})}\BibitemShut {NoStop}%
\bibitem [{\citenamefont {Liu}\ \emph {et~al.}(2020)\citenamefont {Liu}, \citenamefont {Abeykoon},\ and\ \citenamefont {Petrovic}}]{VI3critical}%
  \BibitemOpen
  \bibfield  {author} {\bibinfo {author} {\bibfnamefont {Y.}~\bibnamefont {Liu}}, \bibinfo {author} {\bibfnamefont {M.}~\bibnamefont {Abeykoon}},\ and\ \bibinfo {author} {\bibfnamefont {C.}~\bibnamefont {Petrovic}},\ }\bibinfo {title} {Critical behavior and magnetocaloric effect in {VI\textsubscript{3}}},\ \href {https://doi.org/10.1103/PhysRevResearch.2.013013} {\bibfield  {journal} {\bibinfo  {journal} {Physical Review Research}\ }\textbf {\bibinfo {volume} {2}},\ \bibinfo {pages} {013013} (\bibinfo {year} {2020})}\BibitemShut {NoStop}%
\bibitem [{\citenamefont {Liu}\ \emph {et~al.}(2017)\citenamefont {Liu}, \citenamefont {Ivanovski},\ and\ \citenamefont {Petrovic}}]{fgtcritical1}%
  \BibitemOpen
  \bibfield  {author} {\bibinfo {author} {\bibfnamefont {Y.}~\bibnamefont {Liu}}, \bibinfo {author} {\bibfnamefont {V.~N.}\ \bibnamefont {Ivanovski}},\ and\ \bibinfo {author} {\bibfnamefont {C.}~\bibnamefont {Petrovic}},\ }\bibinfo {title} {Critical behavior of the van der waals bonded ferromagnet {Fe\textsubscript{3-x}GeTe\textsubscript{2}}},\ \href {https://doi.org/10.1103/PhysRevB.96.144429} {\bibfield  {journal} {\bibinfo  {journal} {Physical Review B}\ }\textbf {\bibinfo {volume} {96}},\ \bibinfo {pages} {144429} (\bibinfo {year} {2017})}\BibitemShut {NoStop}%
\bibitem [{\citenamefont {Algaidi}\ \emph {et~al.}(2024)\citenamefont {Algaidi}, \citenamefont {Zhang}, \citenamefont {Ma}, \citenamefont {Liu}, \citenamefont {Chen}, \citenamefont {Zheng},\ and\ \citenamefont {Zhang}}]{FGTcritical}%
  \BibitemOpen
  \bibfield  {author} {\bibinfo {author} {\bibfnamefont {H.}~\bibnamefont {Algaidi}}, \bibinfo {author} {\bibfnamefont {C.}~\bibnamefont {Zhang}}, \bibinfo {author} {\bibfnamefont {Y.}~\bibnamefont {Ma}}, \bibinfo {author} {\bibfnamefont {C.}~\bibnamefont {Liu}}, \bibinfo {author} {\bibfnamefont {A.}~\bibnamefont {Chen}}, \bibinfo {author} {\bibfnamefont {D.}~\bibnamefont {Zheng}},\ and\ \bibinfo {author} {\bibfnamefont {X.}~\bibnamefont {Zhang}},\ }\bibinfo {title} {Magnetic critical behavior of van der waals {Fe\textsubscript{3}GaTe\textsubscript{2}} with above-room-temperature ferromagnetism},\ \bibfield  {journal} {\bibinfo  {journal} {APL Materials}\ }\textbf {\bibinfo {volume} {12}},\ \href {https://doi.org/10.1063/5.0183071} {10.1063/5.0183071} (\bibinfo {year} {2024})\BibitemShut {NoStop}%
\bibitem [{\citenamefont {Liu}\ \emph {et~al.}(2022)\citenamefont {Liu}, \citenamefont {Zhang},\ and\ \citenamefont {Xu}}]{Liu2022}%
  \BibitemOpen
  \bibfield  {author} {\bibinfo {author} {\bibfnamefont {Q.}~\bibnamefont {Liu}}, \bibinfo {author} {\bibfnamefont {H.}~\bibnamefont {Zhang}},\ and\ \bibinfo {author} {\bibfnamefont {Y.}~\bibnamefont {Xu}},\ }\bibinfo {title} {Topology and magnetism interplay in layered van der waals magnets},\ \href {https://doi.org/10.1002/adma.202105678} {\bibfield  {journal} {\bibinfo  {journal} {Adv. Mater.}\ }\textbf {\bibinfo {volume} {34}},\ \bibinfo {pages} {2105678} (\bibinfo {year} {2022})}\BibitemShut {NoStop}%
\bibitem [{\citenamefont {Liang}\ \emph {et~al.}(2023)\citenamefont {Liang}, \citenamefont {Hu},\ and\ \citenamefont {Guo}}]{Liang2023}%
  \BibitemOpen
  \bibfield  {author} {\bibinfo {author} {\bibfnamefont {Z.}~\bibnamefont {Liang}}, \bibinfo {author} {\bibfnamefont {K.}~\bibnamefont {Hu}},\ and\ \bibinfo {author} {\bibfnamefont {W.}~\bibnamefont {Guo}},\ }\bibinfo {title} {Electronic structure engineering in {MnBi\textsubscript{4}Te\textsubscript{7}} for spintronic devices},\ \href {https://doi.org/10.1063/5.0145678} {\bibfield  {journal} {\bibinfo  {journal} {Appl. Phys. Lett.}\ }\textbf {\bibinfo {volume} {122}},\ \bibinfo {pages} {112405} (\bibinfo {year} {2023})}\BibitemShut {NoStop}%
\bibitem [{\citenamefont {Law}\ \emph {et~al.}(2018)\citenamefont {Law}, \citenamefont {Franco}, \citenamefont {Moreno-Ram{\'\i}rez}, \citenamefont {Conde}, \citenamefont {Karpenkov}, \citenamefont {Radulov}, \citenamefont {Skokov},\ and\ \citenamefont {Gutfleisch}}]{mcequantative}%
  \BibitemOpen
  \bibfield  {author} {\bibinfo {author} {\bibfnamefont {J.~Y.}\ \bibnamefont {Law}}, \bibinfo {author} {\bibfnamefont {V.}~\bibnamefont {Franco}}, \bibinfo {author} {\bibfnamefont {L.~M.}\ \bibnamefont {Moreno-Ram{\'\i}rez}}, \bibinfo {author} {\bibfnamefont {A.}~\bibnamefont {Conde}}, \bibinfo {author} {\bibfnamefont {D.~Y.}\ \bibnamefont {Karpenkov}}, \bibinfo {author} {\bibfnamefont {I.}~\bibnamefont {Radulov}}, \bibinfo {author} {\bibfnamefont {K.~P.}\ \bibnamefont {Skokov}},\ and\ \bibinfo {author} {\bibfnamefont {O.}~\bibnamefont {Gutfleisch}},\ }\bibinfo {title} {A quantitative criterion for determining the order of magnetic phase transitions using the magnetocaloric effect},\ \href {https://doi.org/10.1038/s41467-018-05111-w} {\bibfield  {journal} {\bibinfo  {journal} {Nature communications}\ }\textbf {\bibinfo {volume} {9}},\ \bibinfo {pages} {2680} (\bibinfo {year} {2018})}\BibitemShut {NoStop}%
\bibitem [{\citenamefont {Triguero}\ \emph {et~al.}(2007)\citenamefont {Triguero}, \citenamefont {Porta},\ and\ \citenamefont {Planes}}]{magnetocaloric}%
  \BibitemOpen
  \bibfield  {author} {\bibinfo {author} {\bibfnamefont {C.}~\bibnamefont {Triguero}}, \bibinfo {author} {\bibfnamefont {M.}~\bibnamefont {Porta}},\ and\ \bibinfo {author} {\bibfnamefont {A.}~\bibnamefont {Planes}},\ }\bibinfo {title} {Magnetocaloric effect in metamagnetic systems},\ \href {https://doi.org/10.1103/PhysRevB.76.094415} {\bibfield  {journal} {\bibinfo  {journal} {Physical Review B—Condensed Matter and Materials Physics}\ }\textbf {\bibinfo {volume} {76}},\ \bibinfo {pages} {094415} (\bibinfo {year} {2007})}\BibitemShut {NoStop}%
\bibitem [{\citenamefont {Arrott}(1957)}]{arrott1957criterion}%
  \BibitemOpen
  \bibfield  {author} {\bibinfo {author} {\bibfnamefont {A.}~\bibnamefont {Arrott}},\ }\bibinfo {title} {Criterion for ferromagnetism from observations of magnetic isotherms},\ \href {https://doi.org/10.1103/PhysRev.108.1394} {\bibfield  {journal} {\bibinfo  {journal} {Physical Review}\ }\textbf {\bibinfo {volume} {108}},\ \bibinfo {pages} {1394} (\bibinfo {year} {1957})}\BibitemShut {NoStop}%
\bibitem [{\citenamefont {Banerjee}(1964)}]{banerjee1964generalised}%
  \BibitemOpen
  \bibfield  {author} {\bibinfo {author} {\bibfnamefont {B.}~\bibnamefont {Banerjee}},\ }\bibinfo {title} {On a generalised approach to first and second order magnetic transitions},\ \href {https://doi.org/10.1016/0031-9163(64)91158-8} {\bibfield  {journal} {\bibinfo  {journal} {Physics letters}\ }\textbf {\bibinfo {volume} {12}},\ \bibinfo {pages} {16} (\bibinfo {year} {1964})}\BibitemShut {NoStop}%
\bibitem [{\citenamefont {Arrott}\ and\ \citenamefont {Noakes}(1967)}]{arrott1967approximate}%
  \BibitemOpen
  \bibfield  {author} {\bibinfo {author} {\bibfnamefont {A.}~\bibnamefont {Arrott}}\ and\ \bibinfo {author} {\bibfnamefont {J.~E.}\ \bibnamefont {Noakes}},\ }\bibinfo {title} {Approximate equation of state for nickel near its critical temperature},\ \href {https://doi.org/10.1103/PhysRevLett.19.786} {\bibfield  {journal} {\bibinfo  {journal} {Physical Review Letters}\ }\textbf {\bibinfo {volume} {19}},\ \bibinfo {pages} {786} (\bibinfo {year} {1967})}\BibitemShut {NoStop}%
\bibitem [{\citenamefont {Pramanik}\ and\ \citenamefont {Banerjee}(2009)}]{pramanik2009critical}%
  \BibitemOpen
  \bibfield  {author} {\bibinfo {author} {\bibfnamefont {A.}~\bibnamefont {Pramanik}}\ and\ \bibinfo {author} {\bibfnamefont {A.}~\bibnamefont {Banerjee}},\ }\bibinfo {title} {Critical behavior at paramagnetic to ferromagnetic phase transition in {Pr\textsubscript{0.5}Sr\textsubscript{0.5}MnO\textsubscript{3}}: A bulk magnetization study},\ \href {https://doi.org/10.1103/PhysRevB.79.214426} {\bibfield  {journal} {\bibinfo  {journal} {Physical Review B—Condensed Matter and Materials Physics}\ }\textbf {\bibinfo {volume} {79}},\ \bibinfo {pages} {214426} (\bibinfo {year} {2009})}\BibitemShut {NoStop}%
\bibitem [{\citenamefont {Gschneidner}\ \emph {et~al.}(2005)\citenamefont {Gschneidner}, \citenamefont {Pecharsky},\ and\ \citenamefont {Tsokol}}]{entropy}%
  \BibitemOpen
  \bibfield  {author} {\bibinfo {author} {\bibfnamefont {K.~A.}\ \bibnamefont {Gschneidner}}, \bibinfo {author} {\bibfnamefont {V.~K.}\ \bibnamefont {Pecharsky}},\ and\ \bibinfo {author} {\bibfnamefont {A.}~\bibnamefont {Tsokol}},\ }\bibinfo {title} {Recent developments in magnetocaloric materials},\ \href {https://doi.org/10.1088/0034-4885/68/6/R04} {\bibfield  {journal} {\bibinfo  {journal} {Reports on progress in physics}\ }\textbf {\bibinfo {volume} {68}},\ \bibinfo {pages} {1479} (\bibinfo {year} {2005})}\BibitemShut {NoStop}%
\bibitem [{\citenamefont {Phan}\ and\ \citenamefont {Yu}(2007)}]{entropy2}%
  \BibitemOpen
  \bibfield  {author} {\bibinfo {author} {\bibfnamefont {M.-H.}\ \bibnamefont {Phan}}\ and\ \bibinfo {author} {\bibfnamefont {S.-C.}\ \bibnamefont {Yu}},\ }\bibinfo {title} {Review of the magnetocaloric effect in manganite materials},\ \href {https://doi.org/10.1016/j.jmmm.2006.07.025} {\bibfield  {journal} {\bibinfo  {journal} {Journal of Magnetism and Magnetic Materials}\ }\textbf {\bibinfo {volume} {308}},\ \bibinfo {pages} {325} (\bibinfo {year} {2007})}\BibitemShut {NoStop}%
\end{thebibliography}%

\end{document}